# Entangled Photon-Electron States and the Number-Phase Minimum Uncertainty States of the Photon Field

**S Varró**
Research Institute for Solid State Physics and Optics, H-1525 Budapest, POBox 49, Hungary

E-mail: varro@mail.kfki.hu

**Abstract**. The exact analytic solutions of the energy eigenvalue equation of the system consisting of a free electron and one mode of the quantized radiation field are used for studying the physical meaning of a class of number-phase minimum uncertainty states. The states of the mode which minimize the uncertainty product of the photon number and the Susskind and Glogower (1964) cosine operator have been obtained by Jackiw (1968). However, these states have so far been remained mere mathematical constructions without any physical significance. It is proved that the most fundamental interaction in quantum electrodynamics – namely the interaction of a free electron with a mode of the quantized radiation field – leads quite naturally to the generation of the mentioned minimum uncertainty states. It is shown that from the entangled photon-electron states developing from a highly excited number state, due to the interaction with a Gaussian electronic wave packet, the minimum uncertainty states of Jackiw's type can be constructed. In the electron's coordinate representation the physical meaning of the expansion coefficients of these states are the joint probability amplitudes of simultaneous detection of an electron and of a definite number of photons. The photon occupation probabilities in these states preserve their functional form as time elapses, but they depend on the location in space-time of the detected electron. An analysis of the entanglement entropies derived from the photon number distribution and from the electron's density operator is given.

## 1. Introduction

Entanglement and non-locality in quantum mechanics have first been discussed by Einstein, Podolsky and Rosen (1935), and their main conclusion was that quantum mechanics is not a "complete theory", because not every "elements of physical reality" have a counterpart in the theory. As Bohm (1951) writes in his book at the beginning of Section 22.15, "Their criticism has, in fact shown to be unjustified [see Bohr (1935)], and based on assumptions concerning the nature of matter which implicitly contradict the quantum theory at the outset." Motivated by the above work of Einstein, Podolsky and Rosen (EPR), Schrödinger (1935a-b-c) presented a detailed study of the conceptual aspects of quantum mechanics. In this series of papers he introduced the famous "Schrödinger cat" and the concept of "entanglement" ("Verschränkung" in Schrödinger's terminology). In his book in Section 22.15, Bohm (1951) analyses the "EPR-paradox" in detail by considering a desintegration of a quantum system (a molecule having initally zero spin angular momentum) consisting of two spin-1/2 atoms, and detemines the correlations of the spin directions observed at spatially separated detectors. The first reliable experiments, proposed by Wheeler (1946) in this contex, were carried out by Wu and Shaknov (1950), in which they measured coincidence counting rates at different relative azimuths of the polarization of two gamma rays, stemming from electron-positron pair annihilation, and detected by two opposing scintillation counters. They found that the counting rates of perpendicular polarization were two times larger then the rates of parallel polarization. In the optical regime, the first experimental realization of the "Einstein-Podolsky-Rosen-Bohm Gedankenexperiment" have been achieved much later by Aspect et al. (1982a-b). They measured the linear-polarization correlation of pairs of photons emitted in a radiative cascade of calcium, and found excellent agreement with the quantum mechanical predictions, and the greatest violation of generalized Bell's inequalities at that time. Concerning Bell's inequalities see e.g. the references in Aspect et al (1982a-b) and Wigner (1970) and the references therein. In the meantime it turned out that entanglement plays a crucial role in the nowadays rapidly developing branches of quantum physics and informatics, namely in quantum information theory (see e.g. Alber et al 2001, Bouwmeester et al 2001 and Stenholm and Suominen 2005) and in quantum computing and quantum communication (see e.g. Williams 1999 and Nielsen and Chuang 2000).

In the above-mentioned examples the entangled particles are of the same sort. In the present paper we shall discuss entanglement between photons and electrons. It will turn out that the entangled photon-electron states, to be constructed in Section 4, have a close connection with the "critical states" introduced by Jackiw (1968), which minimize a number-phase uncertainty product of the photon field. That is why, concerning the problem of the phase operator of a mode of the quantized electromagnetic radiation, we think that, for the sake of completeness of the present paper, it is instructive to give a brief summary of the basic references dealing with this subject.

In his path-breaking paper on the quantum theory of emission and absorption of radiation Dirac (1927) introduced the photon absorption and emission operators in the form $b_r = e^{-i\theta_r/h} N_r^{1/2}$ and $b_r^* = N_r^{1/2} e^{i\theta_r/h}$, respectively, where, in his notation $r$ is the mode index, $h$ is Planck's constant divided by $2\pi$ and $*$ denotes hermitian conjugation. The number (action) operators $N_r$ and the canonically conjugate angle operators $\theta_r$ are assumed to satisfy the Heisenberg commutation relation, and, as a consequence, $b_r b_r^* - b_r^* b_r = 1$. We note here that in the present paper we shall use the following notations for one mode: $b \to A$, $b^* \to A^+$, thus $[A, A^+] \equiv AA^+ - A^+ A = 1$. The "polar decompositions" used by Dirac is replaced by the relations $A = EN^{1/2}$ and $A^+ = N^{1/2} E^+$, as will be discussed in more details in Section 2. In the same year when Dirac's mentioned paper appeared, London (1927) published his study on the angle variables and canonical transformations in quantum mechanics. He proved that though the ladder operators $E$ and $E^+$ have a well-defined matrix representation, they cannot be expressed as an exponential of the form $e^{\pm i\Phi}$, where $\Phi$ would be a hermitian matrix. It is sure that Dirac was aware of this discrepancy. According to Jordan (1927), in a conversation with him, Dirac remarked that the possibility to derive many correct results by using the *formal* relation $[N, \Phi] = i$ comes from that, the correct relation $[E, N] = E$ has been implicitly used, in fact, instead of the former one, in all the derivations of the results. Formally, the correct relations $ENE^+ = N + 1$ and $E^+ NE = N - 1$ can also be reproduced by assuming $E = e^{i\Phi}$ and $E^+ = e^{-i\Phi}$ with a hermitian $\Phi$, satisfying the commutation relation $[N, \Phi] = i$. At this point let us note that the above-discussed problem of the quantum phase variable does not show up in the case of quantization of the canonically conjugate pair angle and orbital momentum of a planar motion (because the spectrum does not terminate at zero angular momentum), as is illustrated in the extensive and thorough study by Kastrup (2006b), appeared recently.

The non-existence of a hermitian phase operator of a harmonic oscillator was rediscovered by Susskind and Glogower (1964). They introduced the hermitian "cosine" and "sine" operators, whose basic properties will be briefly summarized in Section 2 of the present paper. In their extensive review paper on phase and angle variables in quantum mechanics Carruters and Nieto (1968) derived a couple of number-phase uncertainty relations by using the cosine and sine operators, and Jackiw (1968) constructed a "critical state" which minimizes one of these uncertainty products. Garrison and Wong (1970) constructed a quantum analogon of the classical periodic phase function (saw-tooth), which satisfies the Heisenberg commutation relation with the number operator on a dense set of the Hilbert space of the oscillator. Moreover, they have constructed the eigenstates of this periodic phase operator. In our opinion this was the first mathematically correct approach toward the solution of the original problem of quantum phase. Paul (1974) has proposed an alternative description of the phase of a microscopic electromagnetic field, and discussed the possibilities of its measurement.

A new impetus was given to the study of the quantum phase problem after the papers of Pegg and Barnett (1989) appeared. They truncated the state space of the harmonic oscillator, and were able to construct a hermitian phase operator on this finite-dimensional Hilbert space. We mention that the possibility of using a finite-dimensional (truncated) Hilbert space in this context has already been discussed by Jordan (1927). The approach of Pegg and Barnett (1989) was refined by Popov and Yarunin (1992). The limit matrix elements of the phase operator in number representation (as we let the dimension of the Hilbert space going to infinity) obtained by these authors have already been presented by Weyl (1931). We note that, seemingly, none of the above authors, publishing their papers since the sixties of the last century, had known about the fundamental early works of London (1926 and 1927). The phase distribution of a highly-squeezed states has been determined by Schleich et al (1989) (where the reference to London's work first appeared in the modern era) by using the quantum phase-space distribution (Wigner function) of the quantized mode (see also the book by Schleich 2001, in particular Chapters 8 and 13). The problem of quantum phase measurements has been discussed by Shapiro and Shepard (1991), partly on the basis of "normalizable phase states". The question of operators of phase has been thoroughly analysed by Bergou and Englert (1991) both from the formal point of view and from the physical point of view. In a series of papers Noh et al (1991, 1992a-b and 1993) have studied both theoretically and experimentally the quantum phase dispersion on the basis of their operationally defined cosine and sine operators. In their scheme these definitions are based on measured photon number counts in an eight-port interferometer. Freyberger and Schleich (1993) have performed an analysis of a similar phase operator along with the experiment by Noh et al (1991) by using radially inegrated phase-space distributions. In this context see also the thoroughly written dissertation by Freyberger (1994), and references therein. In the meantime an ample literature has been accumulated concerning the quantum phase problem. For further reading and references we refer the reader to the topical issue of Physica Scripta, edited by Schleich and Barnett (1993), in which also some historical aspects are summarized by Nieto (1993). See also the critical review by Lynch (1995) and the book by Peřinová et al (1999) on the description of phase in optics. Concerning the recent developments of the concept of quantum phase of a linear oscillator, see the thorough group theoretical studies by Kastrup (2003, 2006a and 2007), in which a genuinely new approach to this problem has been worked out.

In the present paper it is proved that the most fundamental interaction in quantum electrodynamics (QED) - namely the interaction of a free electron with a mode of the quantized radiation field - leads quite naturally to the generation of the above-mentioned number-phase minimum uncertainty states. We emphsize that here we are merely dealing with non-relativistic quantum mechanics, where the interaction of the electron with the quantized mode is represented by the minimal coupling term between a free charged particle and an oscillator. The analysis to be presented here is restricted to the study of the interaction of one Schrödinger elctron with one quantized mode of the radiation field. To neglect the interaction with other modes is justified by that we assume a very highly occupied single mode. Thus, in fact, we are not using complete field operators used in the very quantum electrodynamics. In Section 2 we briefly summarize the the basic properties of the Susskind and Glogower (1964) "cosine" and "sine" operators, and we give the associated number-phase uncertainty relations and present the "critical state" found by Jackiw (1968), which minimizes one of the uncertainty products. In Section 3 we present the exact stationary solutions of the photon-electron system, in which the interaction is taken into account up to infinite order. In Section 4 we shall construct the entangled photon-electron states on the basis of these stationary states. It will be shown that the entangled photon-electron states developing from a highly excited number state due to the interaction with a Gaussian electronic wave packet have the same functional form as the "critical states" derived by Jackiw (1968). In Section 5 we derive the reduced density operators of the photon and of the electron. On the basis of these reduced density operators various entanglement entropies are calculated. In Section 6 a short summary closes our paper. The mathematical details of the derivation of our results are presented in the Appendices A and B.

**2. The number-phase minimum uncertainty states of Jackiw**
The number-phase uncertainty product (in contrast to the usual Heisenberg uncertainty products, which are valid e.g. for the variances of the *Cartesian* components of the momentum and position of a particle)
$$(\Delta N)^2 (\Delta \Phi)^2 \geq 1/4 \quad (?) \tag{1}$$
cannot have a well defined mathematical meaning for a generic state of a quantized mode of the electromagnetic radiation. This is because $\Phi$ itself cannot be represented by a matrix (or operator), as London (1927) has already shown long ago. Equation (1) would be valid if there would exist a Heisenberg commutation relation $[N, \Phi] = i$ for the number operator $N$ and for the phase operator $\Phi$, which is not the case here. That is the reason for why Carruthers and Nieto (1968) proposed other uncertainty products given in terms of the $C$ ("cosine") and $S$ ("sine") operators introduced by Susskind and Glogower (1964),

$$C \equiv (E + E^+)/2 \ , \ S \equiv (E - E^+)/2i \tag{2}$$

which are well-defined operators. Here $E$ is the so-called "exponential phase operator" defined by the "*polar decomposition* of the photon absorption operator $A$" (which would be the quantum analogon of the polar decomposition of a complex number, $z = e^{i\varphi}\sqrt{z^*z}$):

$$A = E\sqrt{N} \ , \ N = A^+A \ , \ E = \sum_{k=0}^{\infty}|k\rangle\langle k+1| \ , \ A^+ = \sqrt{N}E^+ \ , \ E^+ = \sum_{k=0}^{\infty}|k+1\rangle\langle k|. \tag{3}$$

We note that the "exponential phase operator", the ladder operator $E$ has been originally introduced by London (1926) and used by Jordan (1927), too. In Eq. (3) $\{|k\rangle, \ k = 0,1,2,...\}$ is the complete orthonormal set of eigenstates of the photon number operator (i.e. $N|k\rangle = k|k\rangle$), serving as a countable basis set of the Hilbert space of the mode under discussion (i.e. $\sum_{k=0}^{\infty}|k\rangle\langle k| = 1$ is the identity operator). Then, owing to the equations $[E,N] = E$ and $[E^+,N] = -E^+$, the following commutation relations can be derived for $N$, $C$, and $S$:

$$[N,C] = -iS \ , \ [N,S] = iC \ , \ [S,C] = P_0/2i \ , \tag{4}$$

where $P_0 \equiv |0\rangle\langle 0|$ is the projector of the vacuum state of the mode, for which $A|0\rangle = 0$. As we see, the "cosine" and the "sine" operators $C$ and $S$, respectively, do not commute, because they cannot be expressed in terms of exponentials of a common (hermitian) operator $\Phi$ in the form $e^{\pm i\Phi}$. The reason for that is the "exponential phase operator" $E$, introduced in Eq. (3), is not unitary but only "half-unitary" (called "partially isometric" in mathematical terminology, see e.g. Riesz and Szőkefalvi-Nagy 1965, Sections 109 and 110). Really, $EE^+ = 1$ holds, but, on the other hand, $E^+E = 1 - P_0$, and, moreover, as a consequence of the half-unitary property of $E$ the sum of the squares the "cosine" and "sine" operators is not equal to unity, $C^2 + S^2 = 1 - P_0/2 \neq 1$. We mention, that for large coherent excitations of the mode, the moments of $C$ and $S$ have a similar form of the moments of the ordinary c-number cosine and sine functions. We have to note here that Kastrup (2006a) has recently raised serious objections against the use of the Susskind and Glogower cosine and sine operators in the description of quantal phase properties of the linear oscillator. On the basis of the analysis presented in Chapter 5 of his paper, he concludes that "the London-Susskind-Glogower operators $\widetilde{C}_k$ and $\widetilde{S}_k$ are *not* appropriate for measuring angle properties of a state!". We would like to emphasize, that in the present study we are not concerned with the question whether the operators $C$ and $S$, defined in Eq. (2), are suitable or not suitable to characterize the quantal phase properties. We merely show that states of essentially the same mathematical structure as that of the "minimizing states" constructed by Jackiw (1968), may be generated in nonperturbative photon-electron interactions in the strong field regime. Thus, we shall not discuss the (questionable or non-existing) physical relevance of $C$ and $S$ themselves in the context of the problem of quantal phase.

The uncertainy products associated to the above commutation relations, Eq. (4), are the following (Carruters and Nieto 1965 and 1968)

$$U_1(\Psi) \equiv (\Delta N)^2 (\Delta C)^2 / \langle S \rangle^2 \geq 1/4 \ , \ U_2(\Psi) \equiv (\Delta N)^2 (\Delta S)^2 / \langle C \rangle^2 \geq 1/4 \ . \tag{5}$$

In the above equations $\Psi$ refers to the state of the quantized mode of the electromagnetic field under discussion, and $(\Delta N)^2$, $(\Delta C)^2$ and $(\Delta S)^2$ are the variances in that state. Jackiw (1968) have constructed a "critical state" which minimizes the first of these uncertainty products, $U_1(\Psi)$,

$$|\Psi\rangle = \sum_{n=0}^{\infty} a_n |n\rangle = \kappa \sum_{n=0}^{\infty} (-i)^n I_{n-\nu}(\gamma) |n\rangle \ , \tag{6}$$

where $I_n$ is a modified Bessel function of first kind of order $n$ (see the definition in Gradshteyn and Ryzhik 2000, formula 8.406.3), and $\kappa$ is a normalization factor. The parameter $\nu \equiv \langle N \rangle$ denotes the mean photon number, $\gamma \equiv \langle S \rangle \neq 0$ and the case, in which $\langle C \rangle \equiv \langle \Psi | C | \Psi \rangle = 0$, has been considered. The expansion coefficients $a_n$ have been determined from the recursion relations $(\nu - n)a_n = (i\gamma/2)(a_{n-1} + a_{n+1})$, coming from the minimizing condition, by taking the subsidiary condition $a_{-1} = 0$ into account. The last requirement (which is equivalent to the equation $I_{-1-\nu}(\gamma) = 0$) forces $\nu$ to satisfy $2s < \nu < 2s+1$, where $s = 0, 1, \cdots$. We have found that this requirement is a consequence of the second theorem of Hurwitz on the zeros of Bessel functions (see Watson 1944, Section 15. 27). The states which allow $U_2(\Psi)$ to reach ¼ can also be constructed, by using the same method. Jackiw (1968) has noted on the states given by Eq. (6) that, "unfortunately these states do not seem to have any physical significance". In the present paper we will show that states of the same structure as that of $|\Psi\rangle$ naturally appear in the non-perturbative analysis of the simplest interaction of QED (namely, the interaction of a free electron with a quantized mode of the electromagnetic radiation). Thus, on the basis of our analysis, we may say that the states to be constructed below, have a fundamental significance.

### 3. Exact energy eigenstates of the interacting photon-electron system

In order to make our paper self-contained, in the present section we briefly summarize the basic steps towards the determination of the exact energy eigenstates of the interacting photon-electron system. We mention that the interaction of electrons with a quantized electromagnetic field within a conducting enclosure has been treated by Smith (1946), but he used perturbation theory, and then rate equations, to treat higher order processes. In his pioneering work on the connection of communication theory and quantum physics, Gabor (1950) also studied a similar system (the transit of electrons in a wave guide), though he used semiclassical pertubation theory and a different geometry.

Let us consider the energy eigenvalue equation of the joint interaction of a quantized mode of the radiation field with a Schrödinger electron. For sake of simplicity, we take for the mode a circularly polarized plane wave in dipole approximation. In this case we do not get squeezing in the stationary states, because the interaction coming from the the $A^2$ term of the Hamiltonian is diagonal. The complete discussions for a Schrödinger electron and for a Dirac electron have been published by Bergou and Varró (1981a-b) and by Bersons (1981) long ago, and have been applied to determine non-perturbatively the cross-sections of multiphoton Bremsstrahlung and multiphoton Compton scattering. Concerning the question of squeezing in photon-electron systems see e.g. Bergou and Varró (1981a), Ben-Aryeh and Mann (1985) and Becker et al (1987). We will consider here only the (simplest) Schrödinger case, and study the interaction with a circularly polarized mode in dipole approximation. The energy eigenvalue equation now reads

$$\left[ \frac{1}{2m}\left( \hat{\vec{p}} + \frac{e}{c}\vec{A} \right)^2 + H_f \right] |\psi_{\vec{p},n_0}\rangle = E_{\vec{p},n_0} |\psi_{\vec{p},n_0}\rangle , \qquad (7)$$

where the vector potential is given as

$$\vec{A} = a(\vec{\varepsilon} A + \vec{\varepsilon}^* A^+) \text{ where } a \equiv (2\pi\hbar c^2/\omega L^3)^{1/2}, \qquad (8)$$

and $\vec{\varepsilon} = (\vec{\varepsilon}_x + i\vec{\varepsilon}_y)/\sqrt{2}$ being the complex polarization vector (for right circular polarization, when the field is assumed to be perpendicular to the z-direction), $\omega$ is the circular frequency of the mode and $L^3$ is the quantization volume. $H_f = \hbar\omega(A^+A + 1/2)$ is the bare field energy. $-e$, $m$ and $c$ have their usual meaning; the elecron's charge and mass, and the velocity of light in vacuum, respectively. $\hbar$ denotes Planck's constant divided by $2\pi$. In Eq. (7) $|\psi_{\vec{p},n_0}\rangle$ are exact stationary states of the interacting photon-electron system characterized by two quantum numbers $\vec{p}$ (the electron's momentum), $n_0$ (a non-negative integer, which, by switching-off the interaction, reduces to the initial photon occupation number). $E_{\vec{p},n_0}$ are the corresponding energy eigenvalues.

The Hamiltonian on the left hand side of Eq. (7) can be rewritten as

$$H = \frac{\hat{\vec{p}}^2}{2m} + \hbar\Omega(A^+A + 1/2) + \frac{ea}{mc}\hat{\vec{p}} \cdot (\vec{\varepsilon} A + \vec{\varepsilon}^* A^+), \quad \Omega \equiv \omega(1 + \omega_p^2/2\omega^2), \quad \omega_p^2 = 4\pi e^2/mL^3 . \qquad (9)$$

Notice that $\omega_p$ is formally nothing else but the plasma frequency for an electron density $1/L^3$. In obtaining Eq. (9) we have taken into account that $\vec{\varepsilon} \cdot \vec{\varepsilon} = 0$, $\vec{\varepsilon}^* \cdot \vec{\varepsilon}^* = 0$ and $\vec{\varepsilon}^* \cdot \vec{\varepsilon} = 1$. The linear interaction term on the right hand side of Eq. (9) can be easily transformed out from the eigenvalue equation, Eq. (7), by applying the following displacement operator with a properly choosen parameter $\sigma$

$$D[\sigma(\vec{p})] = \exp[\sigma^*(\vec{p})A - \sigma(\vec{p})A^+] \quad \text{with} \quad \sigma(\vec{p}) = -(ea/mc\hbar\Omega)\vec{p} \cdot \vec{\varepsilon}^*. \tag{10}$$

We note that the displacement operators of the form displayed by Eq. (10) have an important role in the quantum theory of optical coherence and coherent states, as was first shown by Glauber (1963a-b) in his path-breaking papers. Such displacement operations were also used much earlier by Bloch and Nordsieck (1937) in their fundamental study of the problem of infrared divergeces in QED, in order to transform out the interaction terms. By applying the displacement operation we receive a transformed Hamiltonian which is diagonal in both the electron and the photon variables, hence its eigensolutions can be written down as simple products of the type $|\vec{p}\rangle|n\rangle$, where $|\vec{p}\rangle$ is a momentum eigenstate of the electron. Accordingly, we obtain the eigensolutions of the original Hamiltonian, Eq. (9), in the form

$$|\psi_{\vec{p},n_0}\rangle = |\vec{p}\rangle D[\sigma(\vec{p})]|n_0\rangle. \tag{11}$$

Equation (11) shows that the stationary states of the photon-electron system are products of momentum eigenstates of the electron and generalized coherent states of the photon. If $n_0 = 0$ then the solutions have the structure $|\vec{p}\rangle|\sigma\rangle$, where $|\sigma\rangle$ is an ordinary coherent state. Thus, one may say that (at least, according to the present very simplified description) the self radiation field of the electron is in a coherent state. The complete stationary solutions (being solutions of the time-dependent Schrödinger equation of the joint system) read

$$|\psi_{\vec{p},n_0}(t)\rangle = |\psi_{\vec{p},n_0}\rangle \exp[-iE_{\vec{p},n_0}t/\hbar], \tag{12}$$

where the energy eigenvalues can be brought to the form

$$E_{\vec{p},n_0} = \frac{p_\perp^2}{2m_\perp} + \frac{p_z^2}{2m} + \hbar\Omega(n_0 + 1/2) \quad \text{with} \quad m_\perp = \frac{1 + \omega_p^2/2\omega^2}{1 - \omega_p^2/2\omega^2} m. \tag{13}$$

In Eq. (13) we have used the transverse components $(p_x, p_y) = p_\perp(\cos\chi, \sin\chi)$ of the electron's momentum. It is interesting to note that the "transverse mass" $m_\perp$ given in the second equation of Eq. (13) can in principle be negative (if $\omega_p^2/2\omega^2 > 1$), thus the total energy of the system can also be negative in certain parameter range, which would mean a sort of "attractive interaction" ("bound states") of the mode and of the electron. On the other hand, according to the definition of the one-electron plasma frquency in Eq. (9), for a large enough quantization volume $L^3$, $\omega_p \ll \omega$, thus $m_\perp$ practically equals to the original bare mass $m$. We shall not discuss this question any further in the present paper. For simplicity, in the following we will always assume that $\omega_p^2/2\omega^2 < 1$, thus the "transverse mass" $m_\perp$ is positive. It is clear that if the ratio $\omega_p^2/2\omega^2$ approaches 1 from below, then $m_\perp$ can be much larger then the bare mass $m$ of the electron. For later convenience we rewrite Eq. (12) in the form

$$|\psi_{\vec{p},n_0}(t)\rangle = |p_z\rangle \exp\left[-\frac{i}{\hbar}\frac{p_z^2}{2m}t - i(n_0 + 1/2)\Omega t\right]|\psi_\perp(t)\rangle, \tag{14}$$

$$|\psi_\perp(t)\rangle \equiv |\vec{p}\rangle \exp\left[-\frac{i}{\hbar}\frac{p^2}{2m_\perp}t\right] D[\sigma(\vec{p})]|n_0\rangle. \tag{15}$$

In order to simplify the notation, in Eq. (15) the symbol $\vec{p} \equiv (p_x, p_y)$ has been used for the transverse momentum of the electron, i.e. $\vec{p} \equiv (p_x, p_y) \equiv p(\cos\chi, \sin\chi) = p_\perp(\cos\chi, \sin\chi)$. We note that, owing to the unitarity of the displacement operators, Eq. (10), the exact solutions given by Eq. (11) form a complete orthogonal set on the product Hilbert space $H_{photon} \otimes H_{electron}$ associated to the interacting photon-electron system. The photon statistics of the generalized coherent state of the type $D[\sigma]|n\rangle$, given on the right hand side of Eq. (11), is governed by the matrix elements

$$c_{k,n} \equiv \langle k|D[\sigma]|n\rangle = \begin{cases} (n!/k!)^{1/2} \sigma^{k-n} L_n^{k-n}(|\sigma|^2) e^{-|\sigma|^2/2}, & (k \geq n) \\ (k!/n!)^{1/2} (-\sigma^*)^{n-k} L_k^{n-k}(|\sigma|^2) e^{-|\sigma|^2/2}, & (0 \leq k < n) \end{cases}, \quad (16)$$

where $L_n^s$ denote generalized Laguerre polynamials (for the definition of them see e.g. Gradshteyn and Ryzhik 2000, formula 8.970.1). To our knowledge, the matrix elements of the type given in Eq. (16), was first published in the work by Bloch and Nordsieck (1937), which we have already quoted before. Later Schwinger (1953) derived such matrix elements in one of his famous series of papers on the theory of quantized fields, and they also appear in his study on the Brownian motion of a quantum oscillator (Schwinger 1961). For some further details see e.g. Bergou and Varró (1981a-b). The expectation value of the photon number $\langle k \rangle$, and its variance can be calculated, on one hand, directly from Eq. (16), or, on the other hand, by using the displacement properties $D^+(\sigma)AD(\sigma) = A + \sigma$ and $D^+(\sigma)A^+D(\sigma) = A^+ + \sigma^*$,

$$\langle k \rangle = \sum_{k=0}^{\infty} |c_{k,n_0}|^2 k = \langle n_0|D^+(\sigma)A^+AD(\sigma)|n_0\rangle = n_0 + |\sigma|^2, \quad \Delta k^2 \equiv \langle k^2 \rangle - \langle k \rangle^2 = (2n_0 + 1)|\sigma|^2. \quad (17)$$

## 4. Entangled photon-electron states

In the present section it is proved that the interaction of a free electron with a mode of the quantized radiation field leads to the generation of the number-phase minimum uncertainty states discussed in Section 2. It is shown that the entangled photon-electron states developing from a highly excited number state due to the interaction with a Gaussian electronic wave packet have the same functional form as the minimum "critical states" found by Jackiw (1968). In the electron's coordinate representation the expansion coefficients of these states are expressed in terms of modified Bessel functions of first kind (as has been shown in Eq. (6)) whose argument now depends on the electron's coordinate . The photon statistics of these states preserve their functional form as time evolves, but the occupation probabilities depend on the spatio-temporal position of the electron's detection. We note that on this subject preliminary results have already long been presented by us (Varró 2000), but we have not published them until now.

According to Eqs. (10), (11) and (14), only the transverse motion of the electron couples to the radiation field, thus the longitudinal motion is merely a free propagation. In the following we shall not discuss any further this longitudinal dynamics, but, rather, we concentrate on the study of the transverse part of the wave packet dynamics, which represents in our case the interaction of the electron and the quantized mode of the radiation field. The entangled photon-electron states developing from a number state due to the interaction with an electronic wave packet have the form

$$|\psi\rangle = \int d^2 p\, g(\vec{p})|\psi_\perp(t)\rangle, \text{ with } g(\vec{p}) \equiv g(p) = (w/\hbar\sqrt{\pi})\exp(-p^2 w^2/2\hbar^2), \quad (18)$$

where $g$ has been specialized to a Gaussian weight function, and $|\psi_\perp(t)\rangle$ was introduced in Eq. (15). In Eq. (18) we have introduced the transverse width $w$ of the electronic wave packet (electron beam). The physical situation to which the state given by Eq. (18) may be associated is the following. Let us imagine that an electron is injected into a cavity at time $t = 0$ through a small hole of with $w$. On the basis of our earlier study of the true initial value problem (Bergou and Varró, 1981a), we expect that the sudden coupling of the electron with the (highly occupied) cavity mode, results, *in essence*, in the formation of the state $|\psi\rangle$ defined by Eq. (18). In the present paper we restrict our analysis to the study of the spatio-temporal evolution of these approximate states (which are entangled already at $t = 0$). Owing to the unitarity of the displacement operator $D$ in Eq. (15), the superposition $|\psi\rangle$ defined by Eq. (18) is a normalized state in the product space of the photon-electron system. In order to have an explicit form of this state, we express it in the electron's coordinate representation, and, at the same time, expand it in terms of the photon number eigenstates

$$|\Xi(\vec{r},t)\rangle \equiv \sum_{k=-n_0}^{\infty} |n_0 + k\rangle\langle n_0 + k|\langle\vec{r}|\psi\rangle, \quad \int d^2 r\langle\Xi(\vec{r},t)|\Xi(\vec{r},t)\rangle = 1 . \quad (19)$$

The summation index in the above equation has been shifted merely for the sake of later convenience. The normalization condition in Eq. (19) follows from the proper normalization $\langle \psi | \psi \rangle = 1$ and from the completeness relations $\sum_{n=0}^{\infty} |n\rangle\langle n| = 1_{photon}$, $\int d^2r |\vec{r}\rangle\langle \vec{r}| = 1_{electron}$, where $1_{photon}$ and $1_{electron}$ denote the unit operators on the Hilbert spaces $H_{photon}$ and $H_{electron}$ of the quantized mode and of the electron, respectively. The scalar products in the first equation of Eq. (19) can be expressed as

$$\langle n_0 + k |\langle \vec{r} | \psi \rangle = \int_0^{\infty} dp\, p\, g(p)$$
$$\times \frac{1}{2\pi\hbar} \exp\left(-\frac{i}{\hbar}\frac{p^2}{2m_\perp}t\right) \int_0^{2\pi} d\chi \exp\left[\frac{i}{\hbar} pr \cos(\chi - \varphi)\right] \langle n_0 + k | D[\sigma(\vec{p})] | n_0 \rangle \quad (20)$$

In Eqs. (19) and (20) $r$ and $\varphi$ denote the radial and angular transverse position of the electron, respectively, i.e. $\vec{r} = r(\cos\varphi, \sin\varphi)$. The physical meaning of the matrix elements given by Eq. (20) is that they are joint probability amplitudes of the simultaneous detection of an electron (at position $\vec{r}$ and instant of time $t$) and of a definite number of photons $n_0 + k$. As is shown in Appendix A, for large values of $n_0$ an asymptotic expression can be calculated for the matrix elements of the displacement operator, Eq. (16), (see equations leading to Eq. (A.15)). We note that the integrals over the electron's momentum in Eq. (20) can be evaluated exactly for an arbitrary (not necessarily a large) value of $n_0$ (see the *exact analytic expression* in Eq. (A.7)), but henceforth, in the present paper, we shall only discuss cases of large $n_0$ values, and use the approximation stemming from Eq. (A.15). After the integration with respect to the azimuth angle $\chi$ in momentum space we obtain

$$\langle n_0 + k |\langle \vec{r} | \psi \rangle = (w/\hbar\sqrt{\pi})(-i)^k e^{-ik\varphi}$$
$$\times \int_0^{\infty} dp\, p \exp(-p^2 w^2/2\hbar^2) \exp\left(-\frac{i}{\hbar}\frac{p^2}{2m_\perp}t\right) J_k\left(\sqrt{2}\frac{eA_0}{mc\hbar\Omega}p\right) J_k(pr/\hbar) + O(n_0^{-3/4}). \quad (21)$$

In Eq. (21) we have introduced the quantity $A_0 = (c/\omega)\sqrt{2\pi\rho\hbar\omega}$, which is formally equal to the amplitude of the classical vector potential $\vec{A}_{cl} = A_0(\vec{\varepsilon}e^{-i\omega t} + \vec{\varepsilon}^* e^{i\omega t})$ associated to the photon density $\rho = n_0/L^3$, if make the identification $u = E_{cl}^2/4\pi = \rho\hbar\omega$. Here $u$ denotes the energy density of the mode, with $\vec{E}_{cl} = -\partial \vec{A}_{cl}/\partial ct = (\omega/c)\sqrt{2}A_0(\vec{\varepsilon}_x \sin\omega t - \vec{\varepsilon}_y \cos\omega t)$ being the electric field strength. According to Eq. (A.18), we obtain from Eq. (21) the limit form in case of high initial occupation numbers,

$$\langle n_0 + k |\langle \vec{r} | \psi \rangle = \frac{1}{w\sqrt{\pi}}\frac{(-i)^k e^{-ik\varphi}}{(1+it/\tau)} \exp\left[-\frac{(\mu\Lambda/w)^2 + (r/w)^2}{2(1+it/\tau)}\right] \cdot I_k\left[\frac{(\mu\Lambda/w)\cdot(r/w)}{(1+it/\tau)}\right] + O(n_0^{-3/4}). \quad (22)$$

where $I_k$ is a modified Bessel function of first kind of order $k$, and

$$\mu \equiv \frac{eA_0\sqrt{2}}{mc^2} \equiv \frac{eF_0}{mc\omega} = 10^{-9}\sqrt{I}/E_{ph}, \quad 1/\tau \equiv \hbar/m_\perp w^2 \approx \hbar/mw^2, \quad \Lambda \equiv c/\Omega \approx \lambdabar \equiv \lambda/2\pi. \quad (23)$$

In Eq. (23) we have defined the "dimensionless intensity parameter" $\mu$, whose numerical value can be express in terms of the intensity $I$ of the mode of the radiation field measured in W/cm$^2$, and of the photon energy $E_{ph}$ measured in eV. We have also introduced the amplitude of the electric field strength $F_0 \equiv (\omega/c)A_0\sqrt{2} = \sqrt{4\pi\rho\hbar\omega}$ and the wavelength $\lambda$ of the radiation. The approximate equalities in Eq. (23) are valid for large $L$. If we let both $n_0$ and $L$ going to infinity, in such a way that the photon density is a fixed parameter, then the last term on the right hand side of Eq. (22) can be supressed, and $\mu$ can formally be associated to a classical electric field of amplitude $F_0$. Then in Eq. (22) $\mu\Lambda \to \mu\lambda/2\pi$ becomes just the amplitude of oscillation of a classical electron under the action of the electric field of the radiation $\vec{E} = F_0(\vec{\varepsilon}_x \sin\omega t - \vec{\varepsilon}_y \cos\omega t)$. This can easily be shown by solving the Newton equations $m\ddot{x} = -eE_x$ and $my = -eE_y$. Thus, the dimensionless quantity $\mu\Lambda/w \approx \mu\lambda/2\pi w$ is the ratio of the amplitude of the classical oscillation of the electron to the initial transverse width at $t = 0$ of the electron packet (electron beam). We emphasize that the above remarks were made simply to outline a rough picture in order to give a physical background of the parameters introduced in Eqs. (21) and (23). Of course, we are not saying that a classical electric field can be associated to an even very highly occupied number state. This can consistently be done by using the Schrödinger-Glauber coherent states (Glauber 1963a-b). Anyway, our preliminary investigations on this latter subject clearly show that parameters of a similar sort naturally appear there, too, thus these parameters are allowed to be used in realistic numerical estimates. The time scale parameter $\tau$ defined in Eq. (23) can be related to the period $T = 2\pi/\omega$ of the radiation field through the "bare time scale parameter" $\tau_0 \equiv mw^2/\hbar$,

$$\tau = (m_\perp/m)\tau_0 = \left[(1+\omega_p^2/2\omega^2)/(1-\omega_p^2/2\omega^2)\right]\tau_0, \quad \omega\tau_0 = (1/2)(2mc^2/\hbar\omega)(2\pi w/\lambda)^2. \tag{24}$$

The "transverse mass" $m_\perp$, defined in Eq. (13), can *in principle* be much larger than the "bare mass" $m$, if $\omega_p^2/2\omega^2$ approaches (from below) 1. Consequently, the transverse spreading of the electronic wave packet can *in principle* be reduced due to the interaction with the electromagnetic radiation. From Eq. (19), by neglecting the term of order $n_0^{-3/4}$ in Eq. (22), we have the following approximate form for $|\Xi(\vec{r},t)\rangle$

$$|\Xi(\vec{r},t)\rangle \rightarrow |\widetilde{\Xi}(\vec{r},t)\rangle \equiv \psi_g(r,t) \sum_{k=-n_0}^{\infty} (-i)^k e^{-ik\varphi} I_k[\gamma(r,t)] \cdot |n_0+k\rangle, \tag{25}$$

where

$$\psi_g(r,t) \equiv \frac{1}{w\sqrt{\pi}} \frac{1}{(1+it/\tau)} \exp\left[-\frac{(\mu\Lambda/w)^2 + (r/w)^2}{2(1+it/\tau)}\right], \quad \gamma(r,t) \equiv \frac{(\mu\Lambda/w)\cdot(r/w)}{(1+it/\tau)}. \tag{26}$$

It can be proved by explicit calculation (see the derivation of Eq. (A.22)) that in the limit $n_0 \to \infty$ (and $L \to \infty$, but $n_0/L^3 = \rho$ fixed), the approximate states $|\widetilde{\Xi}(\vec{r},t)\rangle$, given by Eq. (25), are also properly normalized, like the exact states in Eq. (19). By using the index transformation $n_0 + k = n$, we obtain an alternative form of Eq. (25),

$$|\widetilde{\Xi}(\vec{r},t)\rangle = \psi_g(r,t)(-i)^{-n_0} e^{in_0\varphi} \sum_{n=0}^{\infty} (-i)^n e^{-in\varphi} I_{n-n_0}[\gamma(r,t)] \cdot |n\rangle. \tag{27}$$

Apart from the factors $e^{-in\varphi}$, for $t=0$, when $\gamma(r,t)$ is real, the "photon part" (the sum with respect to $n$) on the right hand side of Eq. (27), has the same functional form as the "number-phase minimum uncertainty states" $|\Psi\rangle$, Eq. (6), derived by Jackiw (1968). Notice that the quantum number $n_0$ (corresponding to the parameter $\nu$ in Jackiw's solution) is an integer number in our case, in contrast to $\nu$, which always have to have a non-vanishing fractional part. The other difference is that the normalization constant $\kappa$ in Eq. (6) is determined by the equation $|\kappa|^2 \sum_{n=0}^{\infty} I_{n-\nu}^2(\gamma) = 1$, but in our case $\|\widetilde{\Xi}(\vec{r},t)\|^2 = |\psi_g(r,t)|^2 \sum_{n=0}^{\infty} |I_{n-n_0}[\gamma(r,t)]|^2 \neq 1$ (where $\|\cdot\|$ means the norm in the Hilbert subspace of the quantized mode). For the "photon part" of the state in Eq. (27) a similar normalization to that of Jackiw's states can be achieved by requiring $|\kappa'(r,t)|^2 \sum_{n=0}^{\infty} |I_{n-n_0}[\gamma(r,t)]|^2 = 1$.

At the end of the present Section we give some numerical illustrations of the spatio-temporal behaviour of the joint probabilities $|\langle n_0+k|\langle\vec{r}|\psi\rangle|^2$ on the basis of the analytic expression Eq. (22) found in the large photon excitation limit. These are the probabilities of those simultaneous events when the electron is detected at position $\vec{r}$ *and* $k$ photons are absorbed or emitted at some position (which need not necessarily be the same as that of the electron, rather, for practical reasons, it should be different). In the numerical examples we will always assume that the wavelength of the quantized electromagnetic radiation is of order of $\lambda \approx 10^{-4} cm$, i.e. the photon energy is of order of $\hbar\omega \approx 1 eV$. In this case the dimensionless intensity parameter $\mu$, introduced in Eq. (23), is simply expressed as $\mu = 10^{-9} I^{1/2}$, where $I$ denotes the intensity of the photon field divided by one Watt per square centimeter. Besides, we shall also assume that the wavelength parameter $\Lambda$, introduced in Eq. (23), to a good approximation, coincides with $\lambda/2\pi$. This means, according to the definition of $\Omega$ in Eq. (9), that the one-electron plasma frequency $\omega_p$ is assumed to be much smaller than $\omega$, the frequency of the optical field. In figure 1 we show the spatio-temporal distribution of the joint probabilities $|\langle n_0 + k|\langle \vec{r}|\psi\rangle|^2$ for some given $k$-values.

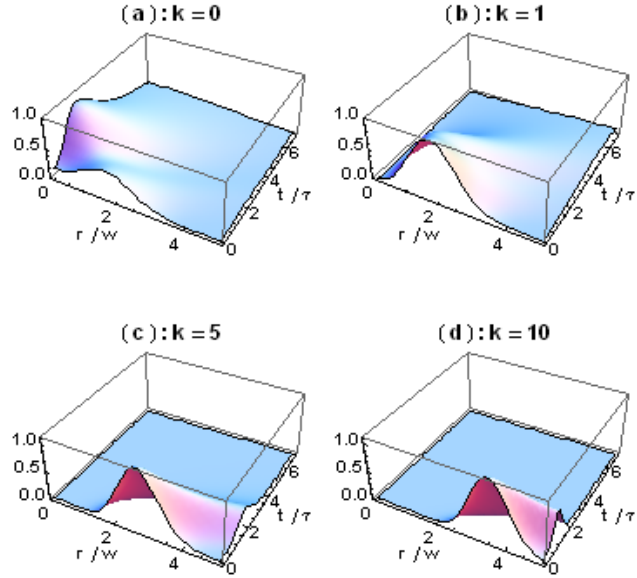

**Figure 1.** Shows the spatio-temporal distribution of the joint probability coming from Eq. (22) or Eq. (25) for different values of the number of emitted excess photons $k$. In each figures we have used a numerical normalization factor, in order to have the maximum values of the vertical coordinate be roughly unity. These factors are the following: (a): 30 for $k=0$, (b): 90 for $k=1$, (c): $2\times 10^4$ for $k=5$, (d): $8\times 10^8$ for $k=25$.

Because of the symmetry of the modified Bessel functions with respect to the change of the sign of their order, $k \to -k$, the same distributions results for photon absorptions. The surfaces in figure 1 illustrate the electron's detection probably at radial position $r$ and at the instant of time $t$, if we know for certainty that $k$ photons have been emitted or absorbed (detected by a spatially separated counter). Here we have taken $(\mu\Lambda/w) = 2$, which corresponds to an intensity $10^{12}$ W/cm². This can be seen from Eq. (23) by assuming that $\lambda/w = 4\pi \times 10^3$. For an optical field $\lambda$ is of order of $10^{-4}$ cm, accordingly $w$ is of order of $10^{-8}$ cm. As is seen, in case of the initial intensity we are considering, the elastic channel ($k=0$) and the one-photon channels ($k=\pm 1$) dominate, and the higher order channels ($|k|>1$) have much less joint probabilities. As is seen in figure 1, for $t=0$ the maxima of the dominant low order joint probabilities ($k=0, \pm 1$) show up at the normalized radial position $\sim \mu\lambda/2\pi w = 2$, which quantity is just the ratio of the amlitude of the electron oscillation to the spatial width of the electronic wave packet. This behaviour can be explained on the basis of the functional form of the position representation of the entangled photon-electron state given by Eqs. (25) and (26).

Our next example, figure 2, illustrates the (joint) photon distribution $|\langle n_0 + k|\langle \vec{r}|\psi\rangle|^2$ around the central large initial photon number $n_0$ for different ratios of $t/\tau$ and $r/w$, i.e. now the spatio-temporal position of the electron detection is a given parameter in each figures. Here $(\mu\Lambda/w) = 4$ is assumed, and $r/w = 10$. The probabilities are normalized to their maximum values, which are $1.10\times 10^{-5}$ in (a), $9.15\times 10^{-5}$ in (b),

$1.97 \times 10^{-4}$ in (c) and $1.86 \times 10^{-4}$ in (d). As the tangent $s$ varies from $0.3$ to $1.5$, the distribution undergoes a qualitative change. The monotonic distribution illustrated by (a) goes over to oscillating distributions, as is shown by (b), (c) and (d).

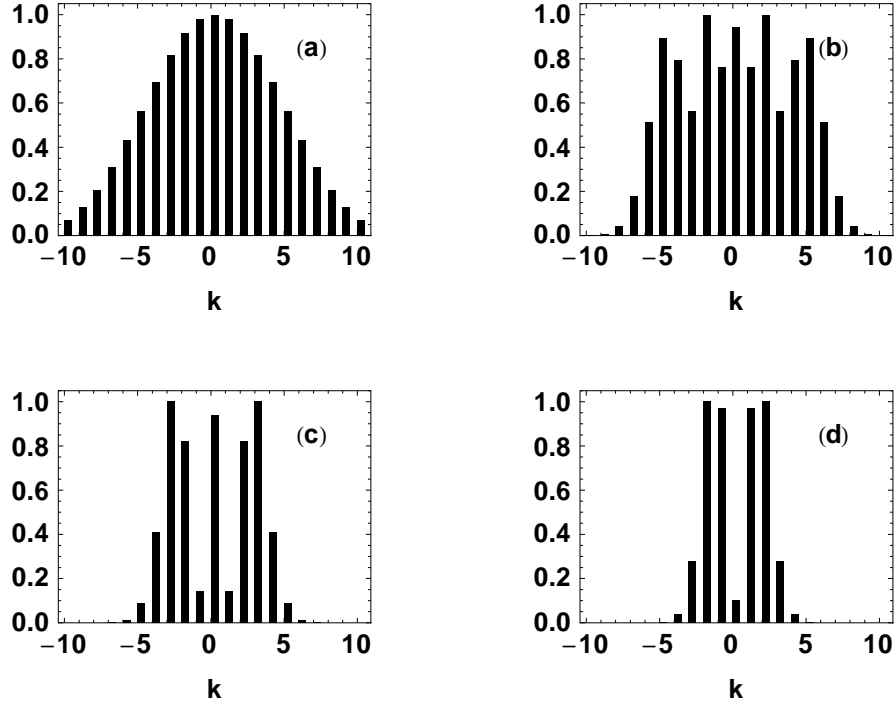

**Figure 2.** Shows the excess photon number distribution around the central large initial photon number $n_0$ for different ratios of $t/\tau$ and $r/w$. This means that the $k$-dependences along the lines $(t/\tau) = s \cdot (r/w)$ on the $r-t$ plane are plotted for different $s$-values. The tangents are: $s = 0.3$ in (a), $s = 0.6$ in (b), $s = 1$ in (c) and $s = 1.5$ in (d).

From figure 2 we can conclude that in certain regions on the $r-t$ plane (where the electron is being detected) the probability distributions of the simultaneous detection of $k$ photons have qualitatively different shapes. It is clear from the functional form of these probabilities, deduced from Eqs. (25) and (26), that in figure 1(a) we see a "*modified* Bessel function behaviour", and on the other hand, in figures 1(b), (c) and (d) we encounter with "*ordinary* Bessel function behaviour". In the first case the distribution has a similar form as the set $\{I_k^2(x)\}$, where $x$ is a real number. In the last three cases the distributions have similar form as $\{J_k^2(x)\}$ for different real values of $x$, and these distributions "oscillate", i.e. there appear local minima and maxima as $k$ varies. We have numerically studied the shapes of the (joint) photon number distributions $|\langle n_0 + k|\langle \vec{r}|\psi\rangle|^2$, and located three regions of the $r-t$ plane where the shapes of the distributions are qualitatively different. The result in a special case is displayed in figure 3. As in figure 2, we assume $(\mu\Lambda/w) = 4$, which corresponds to an intensity $2 \times 10^{12} W/cm^2$ and $\lambda/w = 4\pi \times 10^3$. The tangents of the lower line and the upper line are $0.4$ and $2.8$, respectively. In this case if an electron detection is taking place in the spatio-temporal ranges $(t/\tau) < 0.4 \times (r/w)$ *or* $(t/\tau) > 2.8 \times (r/w)$, then the photon number distributions are one-peaked "monotonic" distributions, like in figure 2(a). On the other hand, in the range defined by the relations $(t/\tau) > 0.4 \times (r/w)$ *and* $(t/\tau) < 2.8 \times (r/w)$ the photon number distributions are "oscillatory", i.e. the joint probability distributions have several local minima and maxima. Of course, the transition from the monotonic regime to the oscillatory regime is not that sharp as the figure would suggest at first glance.

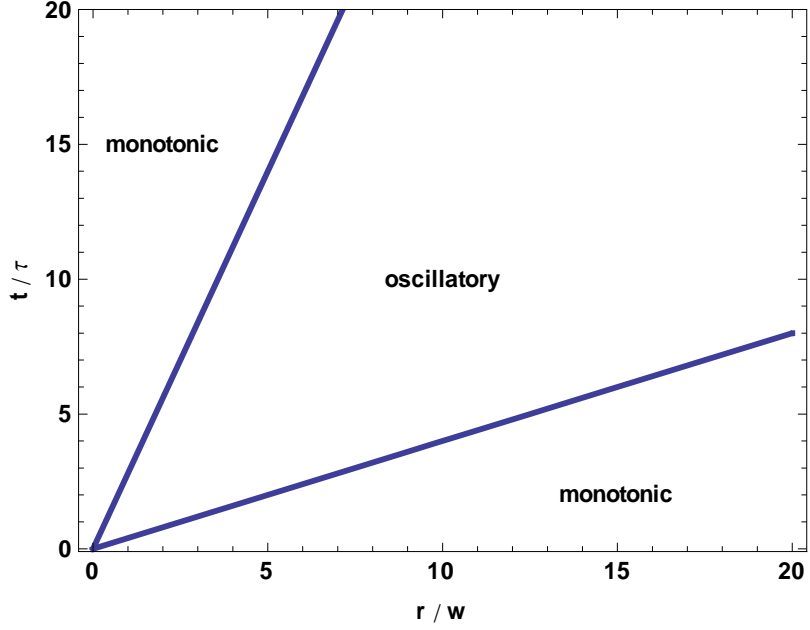

**Figure 3.** Shows schematically the space-time regions where the shapes of the joint probability distributions are qualitatively different.

## 5. Reduced density operators and entanglement entropies

Let us first calculate the density operator $\hat{P}$ of the quantized mode associated to the entangled state $|\psi\rangle$ introduced in Eq. (18). By taking the partial trace (denoted below by $Tr'$) of the dyad $|\psi\rangle\langle\psi|$ with respect to the electron variables, we have

$$\hat{P} \equiv Tr'\{|\psi\rangle\langle\psi|\} = \int d^2 p' \langle \vec{p}'|\psi\rangle\langle\psi|\vec{p}'\rangle = \sum_{k=-n_0}^{\infty} \sum_{l=-n_0}^{\infty} |n_0+k\rangle\langle n_0+l| \int d^2 p \, |g(\vec{p})|^2 \qquad (28)$$
$$\times \langle n_0+k|D[\sigma(\vec{p})]|n_0\rangle \langle n_0+l|D[\sigma(\vec{p})]|n_0\rangle^*$$

In obtaining Eq. (28), the orthogonality of the transverse momentum eigenstates has been used, $\langle \vec{p}|\vec{p}'\rangle = \delta_2(\vec{p}-\vec{p}')$. As is shown in Appendix B, the integral on the right hand side of Eq. (28) can be analytically evaluated, yielding the *exact photon number distribution* given by Eq. (B.3). In the following we shall not discuss this general distribution, but rather, we shall study the case of high initial photon excitations. For large values of $n_0$, the reduced density operator $\hat{P}$ can be brought to the form (see the derivation leading to Eq. (B.7))

$$\hat{P} = \sum_{k=-n_0}^{\infty} |n_0+k\rangle p_k \langle n_0+k| + O(n_0^{-3/4}), \quad p_k \equiv I_k(q)e^{-q}, \qquad (29)$$

where

$$q \equiv (1/2)(\mu\Lambda/w)^2 \approx (1/2)\mu^2(\lambda/2\pi w)^2, \qquad (30)$$

and the quantities $\mu$ and $\Lambda$ have already been defined in Eq. (23). As is proved in Appendix B, the set of weights $\{p_k\}$ is properly normalized, i.e. $\sum_{k=-\infty}^{\infty} p_k = 1$. Owing to the property $I_{-k}(z) = I_k(z)$, the distribution given in Eq. (29) is symmetric to $k=0$, which means that the weights of $k$-photon absorptions are the same as that of $k$-photon emissions. We note that, in fact, the set $\{p_k\}$ governs the *true photon number distribution*, rather than the expansion coefficients of $|\Xi(\vec{r},t)\rangle$ (obtained from Eq. (22), and used in Eqs. (25) and (27)). These latter expansion coefficients are *joint probability amplitudes* of detecting an electron at a position $\vec{r}$, at an instant of time $t$, and, at the same time, detecting $n_0 + k$ photons.

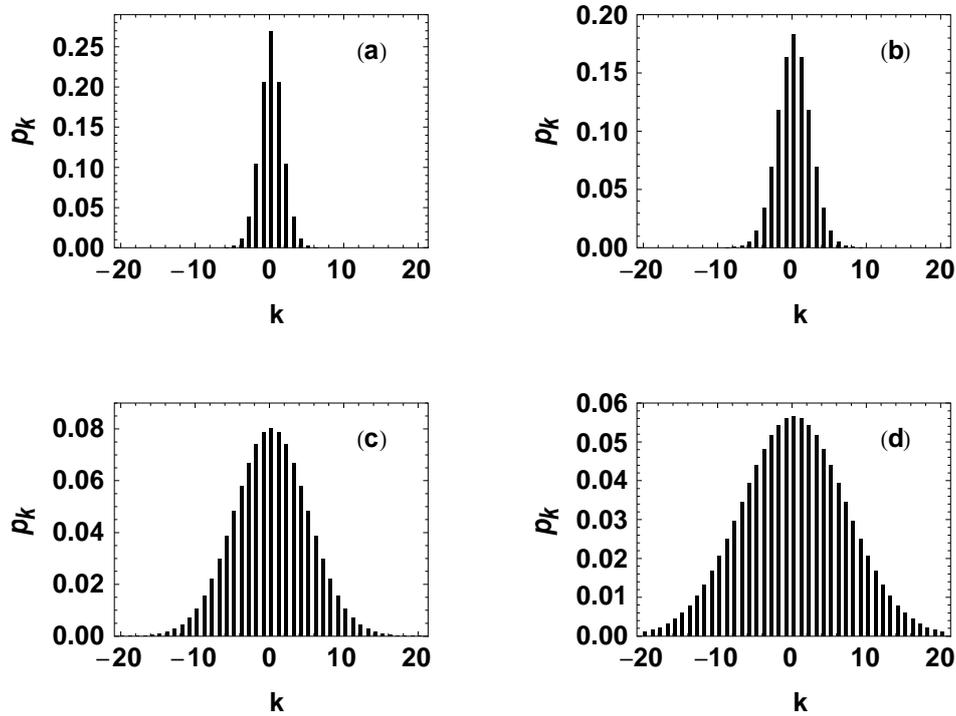

**Figure 4.** Shows the true photon number distribution $\{p_k\}$ (derived from the reduced density operator, and given by Eq. (29)) for four $q$ (intensity) values, namely for $q = 2.5$ in (a), $q = 5$ in (b), $q = 25$ in (c) and $q = 50$ in (d).

If we assume $\lambda/w = 4\pi \times 10^3$, like in figure 1, then these $q$ values used in obtaining figure 4 correspond to intensities of $1.25 \times 10^{12} \ W/cm^2$, $2.5 \times 10^{12} \ W/cm^2$, $1.25 \times 10^{13} \ W/cm^2$ and $2.5 \times 10^{13} \ W/cm^2$, respectively. The terminology "true photon number distribution" we are using for $\{p_k\}$ can be justified by that this set is built up from the (diagonal) elements of the density operator of the photon field, Eq. (29), which, of course, does not contain electron variables, since these latter ones have been traced out. In figure 4 it is clearly seen that as the intensity is increasing the higher order absoption and induced emission events become more and more dominant, and the widths of the distributions are becoming larger and larger. Not an unexpected result. Let us note that the results based on our present analysis do not contradict to the famous statement according to which "a free electron cannot absorb or emit a photon". This statement, which can be found in any of the basic texts on QED, relies on perturbation theory of the S-matrix approach dealing with asymptotic incoming and outgoing plane waves representing the electrons and the photons. The interaction of the electron with a strong laser *beam* is, in fact a many-body interaction, in the sense that the beam can be considered as a superposition of plane electromagnetic waves propagating in different directions, and taking part in high-order induced processes. This question have long been discussed e.g. by Bergou et al (1983), who used a relativistic semiclassical description. The study of such more general problems is out of the scope of the present paper. Here we are using a very simplified scheme (non-relativistic description of the electron, restriction to one mode interactions, dipole approximation, which are, on the other hand, well justified in the range of parameters taken in our numerical examples below). Our goal here is merely to show some basic characteristics of the entangled photon-electron systems.

The von Neumann entropy, $S_{photon}$, associated to the distribution $\{p_k\}$, can be considered as one of the natural measures of the degree of entanglement of the photon-electron system. By using Eq. (29), we obtain

$$S_{photon}[\hat{P}] \equiv -Tr[\hat{P}\log\hat{P}]$$
$$= S_{photon}[\{p_k\}] \equiv -\sum_{k=-\infty}^{\infty} p_k \log p_k = q - \left\{I_0(q)\log[I_0(q)] + 2\sum_{k=1}^{\infty} I_k(q)\log[I_k(q)]\right\}\exp(-q), \quad (31)$$

where $q$ has been defined in Eq. (30). According to Eq. (31) the entropy of the quantized radiation field does not depend on time. This is because the entangled photon-electron state introduced in Eq. (18) in a sense is a *stationary state*, though it contains explicitely the time variable in a complicated manner, as is shown by its analytic form given by Eqs. (25) and (26). The state $|\psi\rangle$, Eq. (18), is *not* a solution of a true initial value problem where we would have assumed an initially non-interacting system (represented by a product state) and switch on the interaction at $t=0$ some way. We leave the study of this latter problem for a separate work in progress (Varró 2007). In Figure 5 we illustrate the intensity dependence of the von Neumann entropy of the photon field. In the parameter range we are considering, the entropy curve, shown in Figure 5 by using log-linear scale, becomes a straight line after the intensity has passed the value $\sim 10^{12} W/cm^2$. This means that the entropy $S_{photon}[\{p_k\}]$ increases logarithmically with the intensity. At zero intensity the entropy vanishes because the interaction of the photon and the electron is negligible in this case (since the foton density is zero).

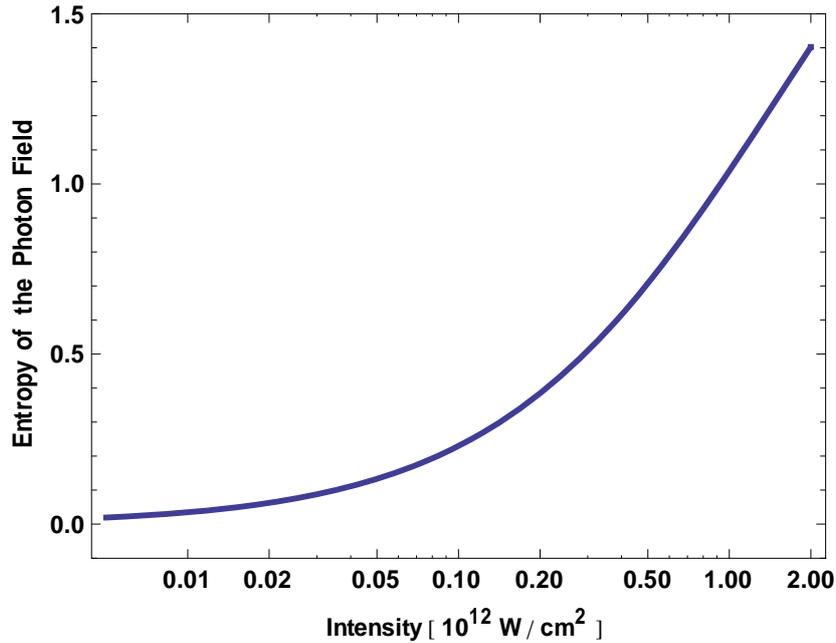

**Figure 5.** Shows the intensity dependence of the von Neumann entropy of the photon field defined by Eq. (31).

In obtaining figure 5 we have assumed that the independent variable $q$ in $S_{photon}[\{p_k\}]$ is expressed numerically as $q = 2\times[I/(W/cm^2)]$. This means, according to Eq. (30), that $\lambda/w = 4\pi \times 10^3$ is assumed, i.e. the wavelength of the radiation field is roughly ten thousands times larger that the initial transverse size of the electronic wave packet. For an optical field we have $\lambda \sim 10^{-4} cm$, so in the case we are considering $w \sim 10^{-8} cm$.

Varró: Entangled_Poton-Electron_Sates_AVarró: Entangled_Poton-Electron_Sates_A                                                             15

Now let us derive the reduced density operator $\hat{P}_e$ of the electron, associated to the entangled state $|\psi\rangle$, which has been introduced in Eq. (18). By taking Eq. (15) into account, the partial trace (denoted by $Tr''$) of $|\psi\rangle\langle\psi|$ with respect to the photon variables reads

$$\hat{P}_e = Tr''\{|\psi\rangle\langle\psi|\} \equiv \sum_{n=0}^{\infty} \langle n|\psi\rangle\langle\psi|n\rangle = \int d^2 p' \int d^2 p'' g(\vec{p}')g^*(\vec{p}'')|\vec{p}'\rangle\langle\vec{p}''|\exp\left[-it(p'^2 - p''^2)/2m_\perp\hbar\right] \quad (32)$$
$$\times \langle n_0 | D^+[\sigma(\vec{p}'')]D[\sigma(\vec{p}')] | n_0 \rangle$$

The matrix elements of $\hat{P}_e$ *in momentum space* can be calculated by using Eqs. (B.10) and (16), yielding

$$P_e(\vec{p}',\vec{p}'') \equiv \langle \vec{p}' | \hat{P}_e | \vec{p}'' \rangle = g(\vec{p}')g^*(\vec{p}'')\exp[-it(p'^2 - p''^2)/2m_\perp\hbar]\exp\{i\,\text{Im}[\sigma(\vec{p}')\sigma^*(\vec{p}'')]\} \quad (33)$$
$$\times L_{n_0}\left(|\sigma(\vec{p}' - \vec{p}'')|^2\right)\exp\left(-|\sigma(\vec{p}' - \vec{p}'')|^2/2\right)$$

where $L_n(x)$ denote Laguerre polynomials of order $n$.

The diagonal matrix elements of $\hat{P}_e$ in momentum space are simply given by the modulus squared of the weight function $g(\vec{p})$ defined in Eq. (18), i.e.

$$P_e(\vec{p},\vec{p}) = |g(\vec{p})|^2 = (w/\hbar)^2 \Pi(\vec{k}), \quad \Pi(\vec{k}) \equiv (1/\pi)\exp(-k^2), \quad \vec{k} \equiv (w/\hbar)\vec{p}. \quad (34)$$

In Eq. (34) we have introduced the dimensionless momentum variable $\vec{k}$ and the density function $\Pi(\vec{k})$. According to Eqs. (B.14) and (B.15), the matrix elements of the reduced density operator $\hat{P}_e$ *in position space* can be espressed as scalar products of the position representation of the entangled photon-electron states introduced in Eq. (19),

$$P_e(\vec{r},\vec{r}';t) = \langle \Xi(\vec{r}',t) | \Xi(\vec{r},t) \rangle = \langle \tilde{\Xi}(\vec{r}',t) | \tilde{\Xi}(\vec{r},t) \rangle + O(n_0^{-3/4}), \quad (35)$$

where $|\tilde{\Xi}(\vec{r},t)\rangle$ has been defined in Eqs. (25) and (26). As is shown in Appendix B, in cases of very high photon excitations (more accurately, in the limit $n_0 \to \infty$) the density function in Eq. (35) becomes

$$P_e(\vec{r},\vec{r}';t) \to \langle \tilde{\Xi}(\vec{r}',t) | \tilde{\Xi}(\vec{r},t) \rangle = (1/w^2)F_e(\vec{x},\vec{x}';t), \quad \text{with} \quad \vec{x} \equiv \vec{r}/w, \quad \vec{x}' \equiv \vec{r}'/w,$$

$$F_e(\vec{x},\vec{x}';t) \equiv \frac{1}{\pi(1+t^2/\tau^2)}\exp\left[-\frac{(\mu\Lambda/w)^2}{(1+t^2/\tau^2)}\right]$$
$$\times \exp\left[-\frac{x'^2 + x^2 + (it/\tau)(x'^2 - x^2)}{2(1+t^2/\tau^2)}\right] \times I_0\left[(\mu\Lambda/w)\frac{x' + x + (it/\tau)(x' - x)}{(1+t^2/\tau^2)}\right]. \quad (36)$$

The diagonal matrix elements of the electron's reduced density operator are determined by the dimensionless density function, which we call *true position distribution of the electron*, since the photon variables have been traced out. We obtain

$$P(\vec{x},t) \equiv F_e(\vec{x},\vec{x};t) = \frac{1}{\pi(1+t^2/\tau^2)}\exp\left[-\frac{(\mu\Lambda/w)^2 + x^2}{(1+t^2/\tau^2)}\right] \times I_0\left[2\frac{(\mu\Lambda/w)\cdot x}{(1+t^2/\tau^2)}\right]. \quad (37)$$

The distribution $P(\vec{x},t)$ is normalized to unity for any instants of time. This can be shown by using a similar procedure applied in the proof of Eq. (A.22).

According to Eq. (29), the density operator of the photon field is diagonal, thus we were able to write down immediately the explicit formula in Eq. (31) for the von Neumann entropy. As is seen from Eqs. (33) and (36), the electron's density operator $\hat{P}_e$, Eq. (32), neither in momentum representation nor in position representation is diagonal. In order to calculate the von Neumann entropy of the electron, first we have to diagonalize $\hat{P}_e$, which, at the moment, seems to us a hopeless task. In order to avoid this difficulty we rather study the so-called *linear entropy H* which has a close connection with the second order *Rényi entropy*. The definition of $H$ reads

$$H \equiv 1 - Tr\hat{\rho}^2 = \exp(H_2) + 1, \quad H_2 \equiv -\log Tr\hat{\rho}^2, \quad (38)$$

where $H_2$ is the second order Rényi entropy, and $\hat{\rho}$ is some generic density operator. The linear entropy has been used by several authors (see e.g. Zurek et al 1993 and Joos et al 2003), because it is much easier to calculate (since we do not need the diagonalization of $\hat{\rho}$), and, on the other hand, it is a good alternative to the von Neumann entropy as a measure of entanglement. Really, $H$ vanishes for a pure state, and it is maximum when the eigenvalues of $\hat{\rho}$ are identical (which is the case of maximum mixing). Another useful quantity to characterize the entanglement in a two particle sytem is the the Schmidt number $K$ (see Nielsen and Chuang 2000) whose definition is $K \equiv [Tr(\hat{\rho}^2)]^{-1}$, where $\hat{\rho}$ denotes the reduced density operator of either one of the two particles. The Schmidt number has been extensively used to charaterize continous-variable entanglement by Fedorov and coworkers (see Fedorov et al 2004, 2005, 2006 and 2007) in their thorough analyses on wave packet dynamics in breakup processes, like ionization of atoms and dissociation of molecules (see, in particular, Fedorov et al 2006, where a unifying overview of rapidly separating systems is presented).

Let us first calculate the linear entropy of the photon field associated to the distribution $\{p_k\}$ given by Eq. (29). The details of the calculation can be found in Appendix B. According to Eq. (B.21) we have

$$H_{photon}[\hat{P}] \equiv 1 - Tr\hat{P}^2 = H_{photon}[\{p_k\}] \equiv 1 - \sum_{k=-\infty}^{\infty} p_k^2 = 1 - e^{-2q} \sum_{k=-\infty}^{\infty} I_k^2(q) = 1 - I_0(2q)e^{-2q}. \tag{39}$$

In order to calculate the linear entropy of the electron, we need first an explicit expression of $\hat{P}_e^2$, which can be obtained from Eq. (B.9) by a straightforward calculation,

$$\hat{P}_e^2 = \int d^2 p_1 \int d^2 p_2 \int d^2 p_3 g(\bar{p}_1) | g(\bar{p}_2) |^2 \, g^*(\bar{p}_3) \exp[-it(p_1^2 - p_3^2)/2m_\perp \hbar]$$
$$\times \langle n_0 | D^+[\sigma(\bar{p}_2)] D[\sigma(\bar{p}_1)] | n_0 \rangle \cdot \langle n_0 | D^+[\sigma(\bar{p}_3)] D[\sigma(\bar{p}_2)] | n_0 \rangle \cdot | \bar{p}_1 \rangle\langle \bar{p}_3 | \tag{40}$$

The trace of $\hat{P}_e^2$ can be calculated analytically, thus we can derive an exact expression for the linear entropy of the electron, as is shown in Appendix B by Eq. (B.28). In the limit $n_0 \to \infty$ (and at the same time the quantization volume $L^3 \to \infty$, such that the photon density $n_0/L^3$ is being kept fixed), according to Eq. (B.29), we obtain

$$Tr\hat{P}_e^2 = \int_0^\infty dx x J_0^2[(\mu\Lambda/w)x]\exp(-x^2/2) = I_0(2q)e^{-2q}, \quad q \equiv \frac{1}{2}(\mu\Lambda/w)^2. \tag{41}$$

By using Eq. (41) and the general definition given by Eq. (38), the linear entropy of the electron becomes
$$H_{electron}[\hat{P}_e] = 1 - I_0(2q)e^{-2q}, \tag{42}$$
which coincides with the linear entropy of the photon field given by Eq. (39),
$$H_{electron}[\hat{P}_e] = H_{photon}[\hat{P}] = 1 - I_0(2q)e^{-2q}. \tag{43}$$

Equation (43) expresses a remarkable consistency in our calculations leading to the analytic results given by Eqs. (39) and (42). Regardless of using the discrete photon number distribution $\{p_k\}$ given by Eq. (29), or using the double integral of the dyads $|\bar{p}'\rangle\langle\bar{p}''|$, parametrized by the two (continuous) momentum variables of the electron in Eq. (32), we end up with the same result for the entanglement entropies. After all, the identity of the two entropies *must* be required for an entangled system consisting of *two* subsystems. In Figure 6 we compare the intensity dependencies of the von Neumann entropy $S_{photon}[\hat{P}]$ and of the (identical) linear entropies $H_{electron}[\hat{P}_e] = H_{photon}[\hat{P}]$ of the electron and of the quantized mode given by Eqs. (31), (42) and (39), respectively. We plotted the curves by using log-linear scale in a larger intensity range than considered in figure 5. As in figure 5, we have assumed that the independent variable $q$ is expressed numerically as $q = 2 \times [I/(W/cm^2)]$, i.e. we have taken $\lambda/w = 4\pi \times 10^3$. The logarithmic increase of the von Neumann entropy with the intensity (curve "S") is clearly seen in the figure. The linear entropy (curve "H") is always smaller than the von Neumann entropy, and increases much slower than the latter one. The increase of each measures of the entanglement by increasing the intensity is after all not an unexpected result, since the interaction of the photons and the free electron is becoming stronger and stronger as the photon density is getting larger.

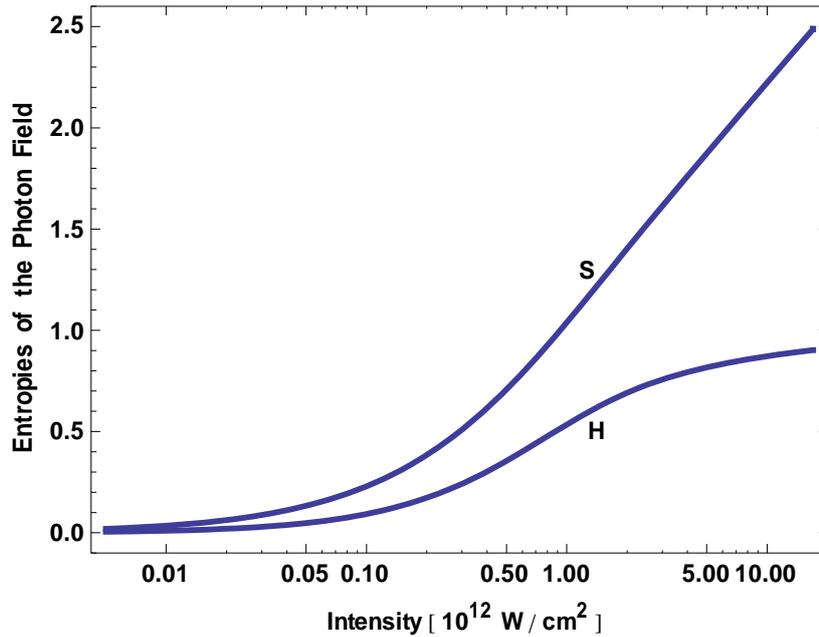

**Figure 6.** Shows a comparison of the intensity dependencies of the von Neumann entropy $S_{photon}[\hat{P}]$ and of the (identical) linear entropies $H_{electron}[\hat{P}_e] = H_{photon}[\hat{P}]$ given by Eqs. (31), (42) and (39), respectively.

## 6. Summary

In the present paper we have discussed entanglement between photons and electrons. We have shown that the entangled photon-electron states introduced by us have a close connection with the critical states introduced by Jackiw (1968), which minimize a number-phase uncertainty product of the photon field. These states are of essentially the same mathematical structure as that of Jackiw's states, and naturally appear in the non-perturbative analysis of the simplest interaction of QED we have considered, namely the interaction of a free electron with a quantized mode of the electromagnetic radiation. On the basis of our analysis we have given a simple interpretation of states of Jackiw's type, thus, we may say that these latter states have a physical significance, rather than being mere mathematical constructions, as they were originally thought of. Besides, we have derived exact analytic expressions for the reduced density operators of the photon field and of the free electron, and determined the von Neumann entropy of the photon, and the linear entropy of the photon and of the electron.

In the introduction we gave a brief historical overview of the development of concepts on entanglement and on the related first basic experiments. Moreover we sketched the most important approaches to the problem of the quantal phase of the linear oscillator (or of a quantized mode of the radiation field). On purpose, we quoted the early references, too, so that the interested reader can keep track of the evolution of concepts on the quantal phase from the very beginning. In Section 2 we have summarized the basic properties of the cosine and sine operators of the quantal phase introduced by Susskind and Glogower (1964), and we presented the critical states constructed by Jackiw (1968), which minimize the uncertainty product of the number operator and the cosine operator. On the basis our earlier work Bergou and Varró (1981a), in Section 3 we determined the exact stationary states of the interacting photon-electron system. These states are simple product states whose photon parts are generalized coherent states, and the electron parts are momentum eigenstates. Section 4 has been devoted to the construction of the entangled photon-electron states which are defined as Gaussian superpositions (with respect to the electron's momentum variable) of the stationary states discussed in Section 3. As we already emphasized, these entangled states defined by Eq. (18) are not solutions of a true initial value problem where we would have assumed an initially non-interacting system (represented by a *bare* product state) and switch on the interaction, say, at $t = 0$ some way. We leave the study of this latter problem for a separate work in progress (Varró 2007). In Appendix A we have given an exact analytic expression for the expansion coefficients of the entangled states (with respect to the number state basis of the photon's Hilbert space and in position representation in the electron's Hilbert space), and in the main text we studied the properties of the associated probability distributions for various parameter values in the large excitation limit. The expansion coefficients of the entangled states, obtained from Eq. (22), and used in Eqs. (25) and (27) are in fact joint probability amplitudes of detecting an electron at some position and at an instant of time, and, at the same time, detecting certain definite number of photons. The basic features of these joint probabilities have been illustrated in Figures 1-2-3. In Section 5 we presented the reduced density operators of the photon field and of the free electron, and with the help of them the true photon number distribution and the electron's momentum and position distributions have been calculated. The exact expressions have been derived in Appendix B, and in the main text we have analysed the characteristics of these distribution in the large excitation limit. As measures of the entanglement, the von Neumann entropy of the photon field and the linear entropies of the photon field and of the electron have also been calculated exactly, and closed analytic forms for them were given in the large excitation limit. We have proved by an explicit calculation that the latter two quantities coincide. Our results are displayed by Figures 4-5-6, which show the true photon number distribution, the intensity dependence of the von Neumann entropy of the photon field and the comparison of the intensity dependence of the linear entropy and of the von Neumann entropy, respectively. Finally we note that it may seem to be a serious restriction to confine our (non-perturbative) study to the analysis of interactions of a free electron with only *one* quantized mode of the radiation field. In reality, of course, the electron interacts with the *whole assembly* of the modes due to e.g. secondary spontaneous emission processes (see for instance the case of Compton scattering). The study of interactions only with one mode can be justified if this mode is in a very highly excited state (as has been mostly assumed in the present paper). In this special case (which, on the other hand, is of great importance in the physics of nonlinear processes taking place in laser-matter interactions) the interactions with the other modes (or with some other third body) can be treated as small perturbations. In this context, see e.g. the works of Bergou and Varró (1981a-b). In order to have an estimate for the magnitude of the excitation degree $n_0$ in realistic laser systems, we can use Eq. (B.34) of Appendix B, which gives a numerical formula for the mean photon occupation number. In Table B 1 we have summarized the numerical values of the parameters we are interested in, for three kinds of laser radiation. It is seen that for intensities managable nowadays, the mean occupation number can be enormously large. Of course, a c-number electric field strength cannot be associated to even a very highly occupied number state in a strict sense. This association can consistently be done, for instance, by using the coherent states of Schrödinger-Glauber's type (Glauber 1963a-b). We plan to present the study of the coherent superpositions of the entangled photon-electron states elsewhere.

**Acknowledgements.** This work has been supported by the Hungarian National Scientific Research Foundation (OTKA), Grant Nos. T48324 and K73728.

**Appendix A**
**Derivation of the explicit form of the entangled photon-electron states**
In the present appendix we show the basic steps leading to the exact analytic form of the matrix elements, Eq. (20),

$$\langle n_0 + k | \langle \vec{r} | \psi \rangle = \int_0^\infty dp\, p g(p)$$

$$\times \frac{1}{2\pi\hbar} \exp\left(-\frac{i}{\hbar}\frac{p^2}{2m_\perp}t\right) \int_0^{2\pi} d\chi \exp\left[\frac{i}{\hbar} pr\cos(\chi - \varphi)\right] \langle n_0 + k | D[\sigma(\vec{p})] | n_0 \rangle \quad\quad\text{(A.1)}$$

and we derive the approximate form, Eq. (22), of them. We prove that the asymptotic states $|\tilde{\Xi}(\vec{r},t)\rangle$, Eq. (25), are properly normalized.

In order to start with, first we give an alternative form of the expansion coefficients of the generalized coherent states, displayed in Eq. (16),

$$c_{n_0+k,n_0} = \langle n_0+k|D[\sigma(\vec{p})]|n_0\rangle = \begin{cases} [n_0!/(n_0+|k|)!]^{1/2}\sigma^{|k|}L_{n_0}^{|k|}(|\sigma|^2)e^{-|\sigma|^2/2}, & (k\geq 0) \\ [(n_0-|k|)!/n_0!]^{1/2}(-\sigma^*)^{|k|}L_{n_0-|k|}^{|k|}(|\sigma|^2)e^{-|\sigma|^2/2}, & (-n_0\leq k<0) \end{cases}, \quad (A.2)$$

where, according to Eq. (10),

$$\sigma(\vec{p}) = -(ea/mc\hbar\Omega)\vec{p}\cdot\vec{\varepsilon}^* = (-e^{-i\chi})bx, \text{ with } b \equiv \frac{ea}{mc\hbar\Omega\sqrt{2}}\frac{\hbar}{w}, \text{ and } x \equiv pw/\hbar. \quad (A.3)$$

The quantities $a$, $\Omega$ and $w$ have been introduced in Eqs. (8), (10) and (18), and the dimensionless variable $x = pw/\hbar$ will be used to calculate the radial integral in Eq. (A.1). The integration with respect to the azimuth angle $\chi$ in momentum space in Eq. (A.1) can be carried out by using the Jacobi-Anger formula for the electron plane wave

$$\exp\left[\frac{i}{\hbar}pr\cos(\chi-\varphi)\right] = \sum_{l=-\infty}^{\infty} i^l J_l(pr/\hbar)e^{il(\chi-\varphi)} \quad (A.4)$$

(see Gradshteyn and Ryzhik 2000, formula 8.551.4), and the elementary relation $\int_0^{2\pi}d\chi e^{in\chi} = 2\pi\delta_{n,0}$. Then, from Eqs. (A.1-4) we obtain

$$\langle n_0+k|\langle\vec{r}|\psi\rangle = (1/w\sqrt{\pi})(-i)^k e^{-ik\varphi}b^s$$
$$\times \int_0^\infty dx\, x^{s+1} e^{-\beta\cdot x^2} J_s(yx) \times \begin{cases} [n_0!/(n_0+s)!]^{1/2} L_{n_0}^s(b^2 x^2), & (k\geq 0) \\ [(n_0-s)!/n_0!]^{1/2} L_{n_0-s}^s(b^2 x^2), & (-n_0\leq k<0) \end{cases}, \quad (A.5)$$

where we have introduced the notation

$$s \equiv |k|, \quad y \equiv r/w, \text{ and } \beta \equiv \frac{1}{2}\left(1+i\frac{t}{\tau}+b^2\right), \text{ with } \tau \equiv \frac{m_\perp w^2}{\hbar}. \quad (A.6)$$

$J_s(z)$ denotes ordinary Bessel function of first kind of order $s$. The radial integral in Eq. (A.5) can be expressed in an analytic form by using the formula 7.421.4 of Gradshteyn and Ryzhik (2000), yielding

$$\langle n_0+k|\langle\vec{r}|\psi\rangle = \frac{(-i)^k e^{-ik\varphi}}{w\sqrt{\pi}2\beta}$$
$$\times \left(\frac{by}{2\beta}\right)^s \exp\left(-\frac{y^2}{4\beta}\right) \begin{cases} \left[\frac{n_0!}{(n_0+s)!}\right]^{1/2}\left(1-\frac{b^2}{\beta}\right)^{n_0} L_{n_0}^s\left[\frac{b^2 y^2}{4\beta(b^2-\beta)}\right] & (k\geq 0) \\ \left[\frac{(n_0-s)!}{n_0!}\right]^{1/2}\left(1-\frac{b^2}{\beta}\right)^{n_0-s} L_{n_0-s}^s\left[\frac{b^2 y^2}{4\beta(b^2-\beta)}\right] & (-n_0\leq k<0) \end{cases}. \quad (A.7)$$

In order to study the asymptotic behaviour of the above exact expression for large $n_0$ values, we use the limit formula (Erdélyi 1953, formula 10.12(36))

$$\lim_{n\to\infty}[n^{-s}L_n^s(z/n)] = z^{-s/2}J_s(2z^{1/2}) \quad (A.8)$$

for given $s$ and $z$ values. Moreover, we take the limit in such a way that, though both $n_0$ and the quantization volume $L^3$ are going to infinity, the photon density $\rho \equiv n_0/L^3$ is considered as a fixed parameter. According to the definitions of $b$ and $\beta$ given by Eqs. (A.3) and (A.6), respectively, we have

$$b = \left(\frac{eA_0\sqrt{2}}{mc^2}\right)\cdot\left(\frac{\Lambda}{w}\right)\cdot\frac{1}{2\sqrt{n_0}} = \frac{1}{\sqrt{n_0}}\frac{1}{2}(\mu\Lambda/w), \text{ where } \Lambda = \frac{c}{\Omega} \to \frac{\lambda}{2\pi},$$

$$A_0 = \left(\frac{c}{\omega}\right)\sqrt{2\pi\left(\frac{n_0}{L^3}\right)\hbar\omega}, \quad \mu = \frac{eA_0\sqrt{2}}{mc^2}, \text{ and } \beta \to \frac{1}{2}\left(1+i\frac{t}{\tau}\right). \quad (A.9)$$

(The physical meaning of the amplitude $A_0$ and the dimensionless intensity parameter $\mu$ is discussed in Section 4 of the main text.) The argument of the Laguerre polynomials and the power expression in front then read, respectively

$$-\frac{b^2 y^2}{4\beta^2} = -\left[\frac{(\mu\Lambda/w)(r/w)}{2(1+it/\tau)}\right]^2 \frac{1}{n_0} \equiv \frac{z}{n_0}, \quad \left(\frac{by}{2\beta}\right)^s = \frac{1}{i^s}\frac{z^{s/2}}{n_0^{s/2}}, \tag{A.10}$$

where the definition of $y$ in Eq. (A.6) has also been taken into account. Thus, on the basis of Eqs. (A.8-10), in the first line on the right hand side of Eq. (A.7) we have

$$\left(\frac{by}{2\beta}\right)^s \left[\frac{n_0!}{(n_0+s)!}\right]^{1/2} L_{n_0}^s\left[-\frac{b^2 y^2}{4\beta^2}\right] = \left[(1+1/n_0)\cdot(1+2/n_0)...(1+s/n_0)\right]^{-1}$$

$$\times (1/i^s) z^{s/2} n_0^{-s} L_{n_0}^s(z/n_0) \to (1/i^s) J_s(2z^{1/2}) = I_s\left[\frac{(\mu\Lambda/w)(r/w)}{(1+it/\tau)}\right] \tag{A.11}$$

where $I_s(\zeta)$ is a modified Bessel function of first kind of order $s$. In obtaining the final result in Eq. (A.11) we have used the relation $J_s(i\zeta) = i^s I_s(\zeta)$, valid for integer $s$ and for an arbitrary complex number $\zeta$ (see Gradshteyn and Ryzhik 2000, formula 8.406.3). The same expression, as in Eq. (A.11), comes out from the second line on the right hand side of Eq. (A.7). The factor $(1-b^2/\beta)^{n_0}$ can be written down in the following alternative form

$$(1-b^2/\beta)^{n_0} = \left[1-\frac{(\mu\Lambda/w)^2}{2(1+it/\tau)}\frac{1}{n_0}\right]^{n_0} \to \exp\left[-\frac{(\mu\Lambda/w)^2}{2(1+it/\tau)}\right], \tag{A.12}$$

where the well-known relation $\lim_{n\to\infty}(1+z/n)^n = e^z$ has been used. From Eqs. (A.7), (A.11) and (A.12), by taking the exponential $\exp(-y^2/4\beta)$ also into account, finally we obtain in the large $n_0$ limit

$$\langle n_0+k|\vec{r}|\psi\rangle = \frac{1}{w\sqrt{\pi}}\frac{(-i)^k e^{-ik\varphi}}{(1+it/\tau)}\exp\left[-\frac{(\mu\Lambda/w)^2+(r/w)^2}{2(1+it/\tau)}\right]\cdot I_k\left[\frac{(\mu\Lambda/w)\cdot(r/w)}{(1+it/\tau)}\right] \quad (n_0\to\infty). \tag{A.13}$$

It is possible to give an alternative (and shorter) derivation of the result, Eq. (A.13), by starting with an asymptotic formula already for the matrix elements Eq. (A.2), instead of using the exact integral Eq. (A.7). In order to do that, let us apply the following formula of Hilb's type (Erdélyi 1953, formula 10.15(2))

$$e^{-x/2} x^{s/2} L_n^s(x) = \frac{\Gamma(n+s+1)}{(\nu/4)^{s/2} n!} J_s[(\nu x)^{1/2}] + O(n^{s/2-3/4}) \tag{A.14}$$

valid for $s > -1$ uniformly in $0 < x \leq K < \infty$, where $\nu \equiv 4n + 2s + 2$. By applying Eq. (A.14), we have

$$\langle n_0+k|D[\sigma(\vec{p})]|n_0\rangle = \left(-e^{-i\chi}\right)^k J_k\left(2\sqrt{n_0}|\sigma(\vec{p})|\right) + O(n_0^{-3/4}). \tag{A.15}$$

The integration with respect to the azimuth angle $\chi$ in momentum space in Eq. (A.1) can again be easily carried out by using the Jacobi-Anger formula, Eq. (A.4). Thus, on the basis of Eq. (A.15) we obtain

$$\langle n_0+k|\vec{r}|\psi\rangle = (w/\hbar\sqrt{\pi})(-i)^k e^{-ik\varphi}$$

$$\times \int_0^\infty dp\, p \exp\left(-p^2 w^2/2\hbar^2\right)\exp\left(-\frac{i}{\hbar}\frac{p^2}{2m_\perp}t\right) J_k\left(\sqrt{2}\frac{eA_0}{mc\hbar\Omega}p\right) J_k(pr/\hbar) + O(n_0^{-3/4}) \tag{A.16}$$

By using Weber's second exponential integral (Watson 1944, formula 13.31(1))

$$\int_0^\infty dx\, x J_k(ax) J_k(bx) e^{-c^2 x^2} = (1/2c^2)\exp\left[-\left(a^2+b^2\right)/4c^2\right] I_k(ab/2c^2) \tag{A.17}$$

(where the condition $|\arg c| < \pi/4$ is to be satisfied), finally we obtain

$$\langle n_0+k|\vec{r}|\psi\rangle = \frac{1}{w\sqrt{\pi}}\frac{(-i)^k e^{-ik\varphi}}{(1+it/\tau)}\exp\left[-\frac{(\mu\Lambda/w)^2+(r/w)^2}{2(1+it/\tau)}\right]\cdot I_k\left[\frac{(\mu\Lambda/w)\cdot(r/w)}{(1+it/\tau)}\right] + O(n_0^{-3/4}), \tag{A.18}$$

which expression is equivalent to that given in Eq. (A.13).

At the end of the present appendix we prove by a direct calculation that the approximate entangled photon-electron states $|\tilde{\Xi}(\vec{r},t)\rangle$, defined in Eq. (25) of Section 4 in the main text, is properly normalized in the limit $n_0 \to \infty$. Really, after the $\varphi$-integration we have

$$\int d^2r \langle \tilde{\Xi}(\vec{r},t)|\tilde{\Xi}(\vec{r},t)\rangle = \frac{2}{w^2(1+t^2/\tau^2)} \exp\left[-\frac{(\mu\Lambda/w)^2}{(1+t^2/\tau^2)}\right]$$
$$\times \int_0^\infty drr \sum_{k=-\infty}^\infty I_k[\gamma^*(r,t)] I_k[\gamma(r,t)] \exp\left[-\frac{(r/w)^2}{(1+t^2/\tau^2)}\right] \quad (A.19)$$

Now, owing to the formulas 8.406.3 and 8.538.1 of Gradshteyn and Ryzhik (2000), $I_k(z) = i^{-k} J_k(iz)$ and

$$\sum_{k=-\infty}^\infty (-1)^k J_k(\xi) J_k(\xi) = J_0(2\xi), \quad (A.20)$$

respectively, we obtain

$$\sum_{k=-\infty}^\infty I_k(\gamma^*) I_k(\gamma) = J_0[i(\gamma^*+\gamma)] = J_0\left[2i\frac{(\mu\Lambda/w)}{\sqrt{1+t^2/\tau^2}} \cdot \frac{(r/w)}{\sqrt{1+t^2/\tau^2}}\right], \quad (A.21)$$

where we have also taken into account the definition of $\gamma(r,t)$ given by Eq. (26). On the basis of Eqs. (A.19) and (A.21), the normalization integral becomes

$$\int d^2r \langle \tilde{\Xi}(\vec{r},t)|\tilde{\Xi}(\vec{r},t)\rangle = 2\exp\left[-\frac{(\mu\Lambda/w)^2}{(1+t^2/\tau^2)}\right] \times \int_0^\infty dxx J_0\left[2i\frac{(\mu\Lambda/w)}{\sqrt{1+t^2/\tau^2}} \cdot x\right] \exp(-x^2) = 1, \quad \forall t, \quad (A.22)$$

thus, the proper normalization "survives" the approximation in the limit $n_0 \to \infty$. In obtaining the result, Eq. (A.22), we have used Weber's first exponential integral

$$\int_0^\infty dxx J_0(ax) e^{-c^2x^2} = \frac{1}{2c^2} \exp\left(-\frac{a^2}{4c^2}\right), \quad (A.23)$$

where $|\arg c| < \pi/4$, and $a$ is an arbitrary complex number (Watson 1944, formula 13.3(1)).

**Appendix B**
**Derivation of the reduced density operators and of the entanglement entropies**

In the present appendix first we calculate the density operator $\hat{P}$ of the quantized mode associated to the entangled state $|\psi\rangle$ introduced in Eq. (18). According to Eq. (28) of Section 4, by taking the partial trace (denoted below by $Tr'$) of the dyad $|\psi\rangle\langle\psi|$ with respect to the electron variables, we have

$$\hat{P} \equiv Tr'\{|\psi\rangle\langle\psi|\} = \int d^2p' \langle\vec{p}'|\psi\rangle\langle\psi|\vec{p}'\rangle = \sum_{k=-n_0}^\infty \sum_{l=-n_0}^\infty |n_0+k\rangle\langle n_0+l| \int d^2p |g(\vec{p})|^2$$
$$\times \langle n_0+k|D[\sigma(\vec{p})]|n_0\rangle \{\langle n_0+l|D[\sigma(\vec{p})]|n_0\rangle\}^* \quad (B.1)$$

By using Eqs. (A.2) and (A.3), after having performed the integration with respect to the azimuth angle $\chi$, from Eq. (B.1) we obtain

$$\hat{P} = \sum_{k=-n_0}^\infty |n_0+k\rangle\langle n_0+k| b^{2s} \int_0^\infty dxx^s \exp[-(1+b^2)x] \times \begin{cases} \frac{n_0!}{(n_0+s)!}\left[L_{n_0}^s(b^2x)\right]^2 & (k \geq 0) \\ \frac{(n_0-s)!}{n_0!}\left[L_{n_0-s}^s(b^2x)\right]^2 & (-n_0 \leq k < 0) \end{cases}, \quad (B.2)$$

where the definitions of $b$ in Eq. (A.3), and the notation $s \equiv |k|$ have been used. The integrals in Eq. (B.2) can be analytically done on the basis of formula 7.414.13 of Gradshteyn and Ryzhik (2000),

$$\hat{P} = \sum_{k=-n_0}^{\infty} |n_0 + k\rangle P_k \langle n_0 + k|, \quad P_k = \frac{(b^2)^{|k|}}{(1+b^2)^{1+|k|}} \begin{cases} P_{n_0}^{(|k|,0)}\left[(1+b^4)/(1-b^4)\right] & (k \geq 0) \\ P_{n_0-|k|}^{(|k|,0)}\left[(1+b^4)/(1-b^4)\right] & (-n_0 \leq k < 0) \end{cases}, \tag{B.3}$$

where $P_n^{(\alpha,\beta)}(x)$ are Jacobi polynomials of order $n$. From Eq. (B.1) it is clear, that, because of the unitarity of the displacement operator $D[\sigma(\bar{p})]$, and, since the profile function $g(p)$ is normalized, the distibution $\{P_k\}$ defined in Eq. (B.3) is normalized to unity. Because $|(1+b^4)/(1-b^4)| > 1$ for both $b > 1$ and $0 < b < 1$, the argument of the Jacobi polynomials in the above equation is out of the interval $[-1, +1]$, in the interior of which all the zeros are located. Thus, the weights $P_k$ cannot take on the zero value. Moreover, out of the interval $[-1, +1]$ this polynomials are all positive (which is, by the way, is to be required from a true probability distribution). In the limit $n_0 \to \infty$, the weights $P_k$ can be approximated with the help of the asymptotic formula (see Erdélyi 1953, formula 10.8(41))

$$\lim_{n \to \infty} n^{-\alpha} P_n^{(\alpha,\beta)}\left(1 - z^2/2n^2\right) = (z/2)^{-\alpha} J_\alpha(z), \tag{B.4}$$

where in our case $n = n_0$ or $n = n_0 - s$, $\alpha = s \equiv |k|$, $\beta = 0$, and from Eq. (A.9)

$$b^2 = (\mu\Lambda/w)^2/2n_0, \quad z = iq, \quad q \equiv (1/2)(\mu\Lambda/w)^2. \tag{B.5}$$

With the help of Eqs. (B.4) and (B.5), the weights in Eq. (B.3) can be approximately expressed in terms of the modified Bessel functions $P_k \approx i^{-|k|} J_{|k|}(iq) = I_k(q)$. However, this approximate distribution is not normalized properly, it only gives the relative photon occupation probabilities. In order to derive a properly normalized approximate distribution in the large $n_0$ limit, directly from Eq. (B.2), we are now proceeding differently as before. By using Hilb's formula, Eq. (A.14), the integral in Eq. (B.2) can be asymtotically expressed as

$$\int_0^\infty dx J_k^2\left[(\mu\Lambda/w)\sqrt{x}\right]e^{-x} = 2\int_0^\infty dx x J_k^2\left[(\mu\Lambda/w)x\right]e^{-x^2} = I_k(q)e^{-q}, \tag{B.6}$$

where we have used Weber's second exponential integral given by Eq. (A.17). Thus, for large values of $n_0$, the reduced density operator $\hat{P}$ can be brought to the form

$$\hat{P} = \sum_{k=-n_0}^{\infty} |n_0 + k\rangle p_k \langle n_0 + k| + O(n_0^{-3/4}),$$

$$p_k \equiv I_k[(1/2)(\mu\Lambda/w)^2]\exp[-(1/2)(\mu\Lambda/w)^2] = I_k(q)e^{-q}, \tag{B.7}$$

where $q \equiv (1/2)(\mu\Lambda/w)^2 \approx (1/2)\mu^2(\lambda/2\pi w)^2$, and the quantities $\mu$ and $\Lambda$ have been already defined in Eq. (A.9) (and in Eq. (23) of Section 4 in the main text). By using the Jacobi-Anger formula, Eq. (A.4), and the relation $I_n(z) = i^{-n} J_n(iz)$, it can be easily proved that $\sum_{n=-\infty}^{\infty} I_n(z) = e^z$, hence the set of weights $\{p_k\}$ is properly normalized, i.e. $\sum_{k=-\infty}^{\infty} p_k = 1$. The von Neumann entropy, $S_{photon}$ associated to the distribution $\{p_k\}$ defined in Eq. (29), can be considered as one of the natural measures of the degree of entanglement in the photon-electron system. By using Eq. (B.7) we obtain

$$S_{photon}[\{p_k\}] = -\sum_{k=-\infty}^{\infty} p_k \log p_k = q - \left\{I_0(q)\log[I_0(q)] + 2\sum_{k=1}^{\infty} I_k(q)\log[I_k(q)]\right\}\exp(-q), \tag{B.8}$$

where the parameter $q$ has been defined after Eq. (B.7).

Now let us derive the reduced density operator $\hat{P}_e$ of the electron, associated to the entangled state $|\psi\rangle$ introduced in Eq. (18). By taking Eq. (15) into account, the partial trace (denoted by $Tr''$) of $|\psi\rangle\langle\psi|$ with respect to the photon variables reads

$$\hat{P}_e = Tr''\{|\psi\rangle\langle\psi|\} \equiv \sum_{n=0}^{\infty} \langle n|\psi\rangle\langle\psi|n\rangle = \int d^2p' \int d^2p'' g(\bar{p}')g^*(\bar{p}'')|\bar{p}'\rangle\langle\bar{p}''|\exp\left[-it(p'^2 - p''^2)/2m_\perp\hbar\right]$$
$$\times \langle n_0|D^+[\sigma(\bar{p}'')]D[\sigma(\bar{p}')]|n_0\rangle$$
(B.9)

The matrix elements of $\hat{P}_e$ in momentum space can be calculated by using the relation

$$D^+[\sigma'']D[\sigma'] = D[\sigma' - \sigma'']\exp[i\,\mathrm{Im}(\sigma''^*\sigma')],$$
(B.10)

and Eq. (16),

$$P_e(\bar{p}', \bar{p}'') \equiv \langle\bar{p}'|\hat{P}_e|\bar{p}''\rangle = g(\bar{p}')g^*(\bar{p}'')\exp[-it(p'^2 - p''^2)/2m_\perp\hbar]\exp\{i\,\mathrm{Im}[\sigma(\bar{p}')\sigma^*(\bar{p}'')]\}$$
$$\times L_{n_0}\left(|\sigma(\bar{p}' - \bar{p}'')|^2\right)\exp\left(-|\sigma(\bar{p}' - \bar{p}'')|^2/2\right)$$
(B.11)

where $L_n(x)$ are Laguerre polynomials of order $n$.

From Eq. (B.11) it is clear that diagonal matrix elements of $\hat{P}_e$ in momentum space are simply given by the modulus squared of the weight function $g(\bar{p})$ defined in Eq. (18), i.e.

$$P_e(\bar{p}, \bar{p}) = |g(\bar{p})|^2 = (w/\hbar)^2 \Pi(\vec{k}), \quad \forall t$$
(B.12)

where we have introduced the dimensionless momentum distribution $\Pi(\vec{k})$,

$$\Pi(\vec{k}) \equiv (1/\pi)\exp(-k^2), \quad \vec{k} \equiv (w/\hbar)\bar{p}.$$
(B.13)

The matrix elements of the reduced density operator $\hat{P}_e$ in position space can be determined by using the identities,

$$P_e(\vec{r}, \vec{r}'; t) \equiv \langle\vec{r}|Tr''\{|\psi\rangle\langle\psi|\}|\vec{r}'\rangle = \sum_{k=-n_0}^{\infty} \langle\vec{r}|\langle n_0 + k|\psi\rangle\langle\psi|n_0 + k\rangle|\vec{r}'\rangle$$
$$= \sum_{k=-n_0}^{\infty}\sum_{k'=-n_0}^{\infty} \langle\psi|\vec{r}'\rangle|n_0 + k'\rangle\langle n_0 + k'|\cdot|n_0 + k\rangle\langle n_0 + k|\vec{r}|\psi\rangle$$
(B.14)

By comparing the factors on the right hand side of Eq. (B.14) with the coordinate representation of the entangled photon-electron state, defined in Eq. (19), we realize that, in fact, the matrix elements are expressed by the following scalar products,

$$P_e(\vec{r}, \vec{r}'; t) = \langle \Xi(\vec{r}', t)|\Xi(\vec{r}, t)\rangle = \langle \tilde{\Xi}(\vec{r}', t)|\tilde{\Xi}(\vec{r}, t)\rangle + O(n_0^{-3/4}),$$
(B.15)

where $|\Xi(\vec{r}, t)\rangle$ has been defined in Eqs. (25) and (26). In case of very high photon excitations (more accurately, in the limit $n_0 \to \infty$) the density function in Eq. (B.15) becomes

$$P_e(\vec{r}, \vec{r}'; t) \to \langle \tilde{\Xi}(\vec{r}', t)|\tilde{\Xi}(\vec{r}, t)\rangle = (1/w^2)F_e(\vec{x}, \vec{x}'; t), \quad \text{with} \quad \vec{x} \equiv \vec{r}/w, \quad \vec{x}' \equiv \vec{r}'/w, \quad \text{and}$$

$$F_e(\vec{x}, \vec{x}'; t) \equiv \frac{1}{\pi(1 + t^2/\tau^2)}\exp\left[-\frac{(\mu\Lambda/w)^2}{(1 + t^2/\tau^2)}\right]$$
$$\times \exp\left[-\frac{x'^2 + x^2 + (it/\tau)(x'^2 - x^2)}{2(1 + t^2/\tau^2)}\right] \times I_0\left[(\mu\Lambda/w)\frac{x' + x + (it/\tau)(x' - x)}{(1 + t^2/\tau^2)}\right].$$
(B.16)

In obtaining Eq. (B.16) we have used the formulas 8.406.3 and 8.538.1 of Gradshteyn and Ryzhik (2000), which were also used in the derivation of Eq. (A.21),

$$I_k(z) = i^{-k}J_k(iz), \quad \sum_{k=-\infty}^{\infty}(-1)^k J_k(z_1)J_k(z_2) = J_k(z_1 + z_2),$$
(B.17)

and the explicit form of $\gamma(r, t)$ given by Eq. (26). The diagonal matrix elements of the electron's reduced density operator are determined by the dimensionless density function

$$P(\vec{x}, t) \equiv F_e(\vec{x}, \vec{x}; t) = \frac{1}{\pi(1 + t^2/\tau^2)}\exp\left[-\frac{(\mu\Lambda/w)^2 + x^2}{(1 + t^2/\tau^2)}\right] \times I_0\left[2\frac{(\mu\Lambda/w)\cdot x}{(1 + t^2/\tau^2)}\right].$$
(B.18)

The density function $P(\vec{x}, t)$ is normalized to unity for any instant of time, as can be shown similarly to the proof of Eq. (A.22).

According to Eq. (B.7), the density operator of the photon field turned out to be diagonal, thus we were able to write down immediately the explicit formula in Eq. (B.8) for the von Neumann entropy,

$$S_{photon}[\hat{P}] \equiv -Tr[\hat{P}\log\hat{P}] = S[\{p_k\}]. \tag{B.19}$$

As we see from Eqs. (B.11) and (B.16), the electron's density operator $\hat{P}_e$, Eq. (B.9), neither in momentum representation nor in position representation is diagonal. In order to calculate the von Neumann entropy of the electron, first we have to diagonalize $\hat{P}_e$, which we have not been able to do by now. In order to avoid this difficulty here we study the so-called *linear entropy* $H$ which has an close connection with the second order *Rényi entropy*. The definition of $H$ reads

$$H \equiv 1 - Tr\hat{\rho}^2 = \exp(H_2) + 1, \quad H_2 \equiv -\log Tr\hat{\rho}^2, \tag{B.20}$$

where $H_2$ is the second order Rényi entropy, and $\hat{\rho}$ is some generic density operator.

Let us first calculate the linear entropy of the photon field associated to the distribution $\{p_k\}$ given by (B.7).

$$H_{photon}[\hat{P}] \equiv 1 - Tr\hat{P}^2 = H_{photon}[\{p_k\}] = 1 - \sum_{k=-\infty}^{\infty} p_k^2 = 1 - e^{-2q}\sum_{k=-\infty}^{\infty} I_k^2(q) = 1 - I_0(2q)e^{-2q}, \tag{B.21}$$

where $q$ has been defined in Eq. (B.5). In deriving Eq. (B.21) we have used the summation theorem of the Bessel functions already given in Eq. (A.20).

In order to calculate the linear entropy of the electron, we need first an explicit expression of $\hat{P}_e^2$, which can be obtained fron Eq. (B.9) by a straightforward calculation,

$$\hat{P}_e^2 = \int d^2 p_1 \int d^2 p_2 \int d^2 p_3 g(\bar{p}_1) |g(\bar{p}_2)|^2 g^*(\bar{p}_3) \exp[-it(p_1^2 - p_3^2)/2m_\perp \hbar] \\ \times \langle n_0|D^+[\sigma(\bar{p}_2)]D[\sigma(\bar{p}_1)]|n_0\rangle \cdot \langle n_0|D^+[\sigma(\bar{p}_3)]D[\sigma(\bar{p}_2)]|n_0\rangle \cdot |\bar{p}_1\rangle\langle\bar{p}_3|. \tag{B.22}$$

The trace of $\hat{P}_e^2$ is the expressed as

$$Tr\hat{P}_e^2 = \int d^2 p_1 \int d^2 p_2 |g(\bar{p}_1)|^2 \cdot |g(\bar{p}_2)|^2 \cdot |\langle n_0|D^+[\sigma(\bar{p}_2)]D[\sigma(\bar{p}_1)]|n_0\rangle|^2. \tag{B.23}$$

By using Eq. (B.10) and (A.2), the above expression can be brought to the form

$$Tr\hat{P}_e^2 = \int d^2 p_1 \int d^2 p_2 |g(\bar{p}_1)|^2 \cdot |g(\bar{p}_2)|^2 \left[L_{n_0}(|\sigma(\bar{p}_1 - \bar{p}_2)|^2)\right]^2 \exp(-|\sigma(\bar{p}_1 - \bar{p}_2)|^2). \tag{B.24}$$

With the help of the following transformation of the integration variables

$$\bar{p}' \equiv \bar{p}_1 - \bar{p}_2, \quad \bar{p}'' \equiv \bar{p}_1 + \bar{p}_2, \quad \text{with} \quad \left|\frac{\partial(p_{1x}, p_{1y}, p_{2x}, p_{2y})}{\partial(p'_x, p'_y, p''_x, p''_y)}\right| = \frac{1}{4}, \tag{B.25}$$

the integrations with respect $\bar{p}'$ and $\bar{p}''$ can be separated,

$$Tr\hat{P}_e^2 = (1/4)(w/\hbar\pi^{1/2})^2 (w/\hbar\pi^{1/2})^2 \int d^2 p'' \exp(-p''^2 w^2/2\hbar^2) \\ \times \int d^2 p' \left[L_{n_0}(|\sigma(\bar{p}')|^2)\right]^2 \exp(-|\sigma(\bar{p}')|^2)\exp(-p'^2 w^2/2\hbar^2), \tag{B.26}$$

where we have used the explicit form $g(\bar{p}) = (w/\hbar\pi^{1/2})\exp(-p^2 w^2/2\hbar^2)$ of the momentum profile function introduced in the main text in Eq. (18). After having performed the $\bar{p}''$-integration, and introducing dimensionless parameters and integration variables we have from Eq. (26)

$$Tr\hat{P}_e^2 = \int dx x \left[L_{n_0}(b^2 x^2)\right]^2 \exp[-(b^2 + 1/2)x^2] = \frac{1}{2b^2}\cdot\left(1 - \frac{1}{2b^2}\right)^{n_0} P_{n_0}\left[\frac{1+(1/2b^2)^2}{1-(1/2b^2)}\right], \tag{B.27}$$

where $P_n(x)$ are Legendre polynomials of order $n$. In obtaining Eq. (B.27) we have used the definitions in Eq. (A.3) and the formula 7.414.2 of Gradshteyn an Ryzhik (2000). According to Eq. (B.27) and the general formula given by Eq. (20), the exact expression for the electron's linear entropy can be writtren as

$$H_{electron}[\hat{P}_e] = 1 - \frac{1}{2b^2}\cdot\left(1 - \frac{1}{2b^2}\right)^{n_0} P_{n_0}\left[\frac{1+(1/2b^2)^2}{1-(1/2b^2)}\right], \quad b \equiv \frac{ea}{mc\hbar\Omega\sqrt{2}}\frac{\hbar}{w} = \frac{1}{\sqrt{n_0}}\frac{1}{2}(\mu\Lambda/w), \tag{B.28}$$

where the definitions in Eq. (A.9) has also been used. In the large $n_0$ limit we can use Hilb's formula, Eq. (A.14), and apply it directly in the integrand in Eq. (B.27). We have

$$Tr\hat{P}_e^2 = \int_0^\infty dx\, x J_0^2[(\mu\Lambda/w)x]\exp(-x^2/2) + O(n_0^{-3/4}) = I_0(2q)e^{-2q} + O(n_0^{-3/4}), \quad q \equiv \frac{1}{2}(\mu\Lambda/w)^2. \quad (B.29)$$

In obtaining the above analytic result we have used Weber's second exponential integral formula displayed already in Eq. (A.17). By using Eq. (B.29), the electron's linear entropy in the large $n_0$ limit turns out to be

$$H_{electron}[\hat{P}_e] = 1 - I_0(2q)e^{-2q}, \quad (B.30)$$

which coincides with the linear entropy of the photon field given by Eq. (B.21),

$$H_{electron}[\hat{P}_e] = H_{photon}[\{p_k\}] = 1 - I_0(2q)e^{-2q}. \quad (B.31)$$

At the end of the present appendix we give an estimate for the average occupation number of the photon field expressed in terms of the amplitude of the electric field strength $F_0$, or, equivalently, in terms of the intensity $I$ of a quasi-monochromatic radiation. In free space the three-dimensional spatial mode density in a frequency interval $(\omega, \omega + \Delta\omega)$ can be expressed as $Z_\omega \Delta\omega$, where $Z_\omega = \omega^2/\pi^2 c^3$ is the spectral mode density. We equate the time average of the energy density $F_0^2/4\pi$ of the radiation with the mode density $Z_\omega \Delta\omega$ times the average occupation number $\bar{n}$ of the modes multiplied by the central photon energy $\hbar\omega$, i.e.

$$Z_\omega \Delta\omega \cdot \bar{n} \cdot \hbar\omega = F_0^2/4\pi \quad \rightarrow \quad \bar{n} = \frac{\pi^2 c^3}{\omega^2} \cdot \frac{1}{\Delta\omega\hbar\omega} \cdot \frac{F_0^2}{4\pi}. \quad (B.32)$$

By introducing dimensionless combinations of the parameters, we obtain from Eq. (B.32)

$$\bar{n} = \frac{\pi}{16}\left(\frac{\hbar c}{e^2}\right)\cdot\left(\frac{2mc^2}{\hbar\omega}\right)^2 \cdot \left(\frac{\omega}{\Delta\omega}\right)\cdot\left(\frac{eF_0}{mc\omega}\right)^2 = \frac{\pi}{16\alpha}\cdot\left(\frac{2mc^2}{\hbar\omega}\right)^2 \cdot \left(\frac{\omega}{\Delta\omega}\right)\cdot \mu^2, \quad (B.33)$$

where $\alpha \equiv e^2/\hbar c \approx 1/137$ denotes the fine structure constant and $\mu \equiv eF_0/mc\omega$ is the dimensionless intensity parameter introduced already in Eq. (23) of Section 4. The second factor on the right hand side of the last equation in Eq. (B.33) is the square of the ratio of the electron-positron pair creation energy $2mc^2 \approx 10^6\, eV$ to the photon energy $\hbar\omega$. For optical fields this factor is of order of $10^{12}$, because in this case $\hbar\omega \approx 1eV$. The third factor $\omega/\Delta\omega$ is the inverse relative bandwidth of the radiation, which, in case of pulsed lasers, is of order of $\tau_{pulse}/T$, where $\tau_{pulse}$ is the pulse duration and $T$ is the period of the central spectral component. On the basis of Eq. (B.33), the numerical value of $\bar{n}$ can be calculated according to the formula

$$\bar{n} = 2.7 \times 10^{-5}(\omega/\Delta\omega)\cdot(I/E_{ph}^4), \quad (B.34)$$

where $I$ denotes the intensity divided by $W/cm^2$, and $E_{ph}$ is the photon energy measured in $eV$-s. In Table B 1 we summarize for three kinds of lasers the numerical values of the photon energy, the inverse bandwidth and the corresponding average photon occupation number, expressed in terms of the dimensionless intensity $I$, on the basis of Eq. (B.34).

|  | *5.2 fs Ti : Sa laser* | *5.2 ps Nd : Glass laser* | *CW He-Ne laser* |
|---|---|---|---|
| $E_{ph}$ | 1.57 | 1.17 | 1.96 |
| $\omega/\Delta\omega$ | 2 | $3\times 10^3$ | $10^8$ |
| $\bar{n}$ | $9\times 10^{-6}\, I$ | $6\times 10^{-2}\, I$ | $2\times 10^3\, I$ |

**Table B.1** Shows for three kinds of lasers the numerical values of the photon energy, the inverse bandwidth and the corresponding average photon occupation number.

**Caption to Figure 1**
Shows the spatio-temporal distribution of the joint probability coming from Eq. (22) or Eq. (25) for different values of the number of emitted excess photons $k$. In each figures we have used a numerical normalization factor, in order to have the maximum values of the vertical coordinate be roughly unity. These factors are the following: (a): 30 for $k=0$, (b): 90 for $k=1$, (c): $2 \times 10^4$ for $k=5$, (d): $8 \times 10^8$ for $k=25$.

**Caption to Figure 2**
Shows the excess photon number distribution around the central large initial photon number $n_0$ for different ratios of $t/\tau$ and $r/w$. This means that the $k$-dependences along the lines $(t/\tau) = s \cdot (r/w)$ on the $r-t$ plane are plotted for different $s$-values. The tangents are: $s=0.3$ in (a), $s=0.6$ in (b), $s=1$ in (c) and $s=1.5$ in (d).

**Caption to Figure 3**
Shows schematically the space-time regions where the shapes of the joint probability distribution are qualitatively different.

**Caption to Figure 4**
Shows the true photon number distribution $\{p_k\}$ (derived from the reduced density operator, and given by Eq. (29)) for four $q$ (intensity) values, namely for $q=2.5$ in (a), $q=5$ in (b), $q=25$ in (c) and $q=50$ in (d).

**Caption to Figure 5**
Shows the intensity dependence of the von Neumann entropy of the photon field defined by Eq. (31).

**Caption to Figure 6**
Shows a comparison of the intensity dependencies of the von Neumann entropy $S_{photon}[\hat{P}]$ and of the (identical) linear entropies $H_{electron}[\hat{P}_e] = H_{photon}[\hat{P}]$ given by Eqs. (31), (42) and (39), respectively.

**Caption to Table B.1**

Shows for three kinds of lasers the numerical values of the photon energy, the inverse bandwidth and the corresponding average photon occupation number.